\documentclass[journal=ancac3,manuscript=article,layout=twocolumn]{achemso}

\usepackage[version=3]{mhchem} 
\usepackage{placeins}
\usepackage{booktabs}
\usepackage{xcolor}

\newcommand*{\citen}[1]{%
  \begingroup
    \romannumeral-`\x 
    \setcitestyle{numbers}%
    \cite{#1}%
  \endgroup   
}


\author{Will T. Kaufhold}
{\email{willtkaufhold@gmail.com}}
\affiliation[Cam]
{Department of Physics, University of Cambridge, JJ Thomson Avenue, Cambridge CB3 0HE, UK}
\alsoaffiliation[Imp]
{Department of Chemistry, Molecular Sciences Research Hub, Imperial College London, London W12 0BZ, UK}

\author{Wolfgang Pfeifer}
\affiliation[Ohio]
{Department of Mechanical and Aerospace Engineering, The Ohio State University, Columbus, OH 43210, USA}
\author{Carlos E. Castro}
\affiliation[Ohio]
{Department of Mechanical and Aerospace Engineering, The Ohio State University, Columbus, OH 43210, USA}

\author{Lorenzo Di Michele}
\email{l.di-michele@imperial.ac.uk}
\affiliation[Imp]
{Department of Chemistry, Molecular Sciences Research Hub, Imperial College London, London W12 0BZ, UK}
\alsoaffiliation[Cam]
{Department of Physics, University of Cambridge, JJ Thomson Avenue, Cambridge CB3 0HE, UK}
\alsoaffiliation[fabricell]
{fabriCELL, Molecular Sciences Research Hub, Imperial College London, London W12 0BZ, UK}

\title[]{Probing the mechanical properties of DNA nanostructures with metadynamics}


\begin{document}


\begin{abstract}
    Molecular dynamics simulations are often used to provide feedback in the design workflow of DNA nanostructures. However, even with coarse-grained models, convergence of distributions from unbiased simulation is slow, limiting applications to equilibrium structural properties. Given the increasing interest in dynamic, reconfigurable, and deformable devices, methods that enable efficient quantification of large ranges of motion, conformational transitions, and mechanical deformation are critically needed. Metadynamics is an automated biasing technique that enables the rapid acquisition of molecular conformational distributions by flattening free energy landscapes. Here  we leveraged this approach to sample the free energy landscapes of DNA nanostructures whose unbiased dynamics are non-ergodic, including bistable Holliday junctions and part of a bistable origami. Taking an origami compliant joint as a case study, we further demonstrate that metadynamics can predict the mechanical response of a full DNA origami device to an applied force, showing good agreement with experiments. Our results establish an efficient framework to study free energy landscapes and force response in DNA nanodevices,  which could be applied for rapid feedback in iterative design workflows and generally facilitate the integration of simulation and experiments. Metadynamics will be particularly useful to guide the design of dynamic devices for nanorobotics, biosensing, or nanomanufacturing applications.
\end{abstract}

\section{Introduction}

In structural DNA nanotechnology, a collection of DNA sequences is chosen to form a desired structure \emph{via} molecular self-assembly \cite{Seeman2005,Ke2018}. Such DNA constructs often have a single well-defined free energy minimum, corresponding to geometries like ribbons \cite{Winfree1998}, tiles \cite{rothemund2006}, square or honeycomb arrangements of helices \cite{Ke2009,Douglas2009b}, or brick-like voxel arrays of short DNA oligonucleotides\cite{Ke2012}. These unimodal structures (\emph{i.e.} having one primary configuration in space) have been translated to applications where structural rigidity is important -- in fiducials for super-resolution microscopy \cite{Raab2018}, as scaffolds to visualize biomolecular processes \cite{Rajendran2014}, or as nanopores for single molecule detection \cite{Bell2012}.\\

With the DNA origami technique, a bacteriophage genome and synthetic oligonucleotides co-assemble to create near-arbitrary shapes \cite{rothemund2006,Ke2009,Douglas2009,Douglas2009b}.  { DNA origami has emerged as a dominant approach in nanoscale structural design, and unlocked the manufacture of nanostructures programmed to perform complex motion \cite{Zhou2015,Huang2020,Ijas2018_referee2}, \emph{e.g.} hinges~\cite{castro2015}, pistons~\cite{Marras2015}, interlocked axles and sliders~\cite{Marras2015,List2016_referee2}, and rotors~\cite{Philip2020_referee2,Kosuri2019_referee2,Enzo2018_referee2}. These deformable elements have formed the basis of stimuli responsive materials \cite{Gerling2015}, sensors \cite{Kuzyk2014}, single-molecule probes\cite{Le2016,Stephanopoulos2021,Kosuri2019_referee2}, drug delivery vectors~\cite{M2012_referee2}, and nanoreactors.~\cite{Grossi2017_referee2}} The motion of origami nanomachines can be constrained to occur along given axes, and configurational distributions can feature multiple stable states separated by energy barriers~\cite{Song2017,Zhou2015}.\\

Molecular modelling has become a key element in the design workflow of DNA nanostructures, with the two most common approaches being finite-element modelling, and Molecular Dynamics (MD). Finite-element frameworks, such as Cando \cite{Castro2011} and SNUPI \cite{Lee2021}, describe DNA helices as elastic rods, and apply continuum mechanics to predict the equilibrium structure and its deformation modes. The latter are however only accurate in describing small deformations, and become poor approximations when the structure deforms significantly, or has multiple stable states.  Additionally, these continuum approaches lack the resolution to describe molecular processes such as formation and dissociation of base pairing and stacking bonds, which may be critical for the behavior of dynamic devices.\\

Conversely, MD infers mechanical properties by explicit simulation of the system's Newtonian dynamics. Atomistic simulations of DNA nanostructures may take weeks to complete \cite{Yoo2013}, motivating the development of coarse-grained models such as the multi resolution DNA (MrDNA) framework \cite{Maffeo2020}, and oxDNA \cite{Snodin2015}.\\

Thanks to its ability to accurately represent nucleotide stacking and base pairing, the oxDNA force field \cite{Ouldridge2011,Snodin2015} has succeeded in replicating various phenomena, including kinking in duplexes \cite{Harrison2015,Harrison2015a} and force-induced unravelling of origami \cite{Engel2018}, and has been applied to predict conformational distributions of origami mechanical elements \cite{Shi2017,Sharma2017}.  As a result, oxDNA is now frequently used as part of iterative nanostructure design workflows\cite{Huang2020,Benson2019}.\\

However, even coarse-grained simulations can be impractically slow~\cite{Sharma2017}, and without \emph{ad hoc} biasing techniques can only sample configurations with free energy within a few $k_BT$ away from the minima. Additionally, trajectories can become trapped in local minima, hindering complete sampling. As a result, coarse-grained simulations are often performed merely to check for mechanical strain or undesired deformations, instead of quantitatively assessing of the range of motion or the forces required for actuation.\\

Various biasing techniques can be used to flatten free energy landscapes and accelerate sampling \cite{Pietrucci2017}. These approaches use fictitious forces along collective variables, which are low dimensional representations of conformational states.  One such method, previously integrated with oxDNA~\cite{Engel2019},  combines steered MD with the use of the Jarzynski equality \cite{Jarzynski1997} to reconstruct free energy landscapes along a 1D reaction coordinate. This method is however unsuitable for acquiring multidimensional landscapes, and its estimates are dominated by unlikely low-work trajectories, resulting in difficult to assess uncertainties \cite{Gore2003}.\\

Shi \emph{et al.}~\cite{Shi2020}, and more recently, Wong \emph{et al.}~\cite{wong2021characterizing} have demonstrated that the integration of umbrella sampling with oxDNA can enable the exploration of 1D and 2D free energy landscapes associated with the deformation of origami, { while this technique had been previously applied to exploring deformations in smaller nanostructures, including duplex bending~\cite{Harrison2015a} and junction flexibility~\cite{Snodin2019}. Umbrella sampling} relies on defining multiple (partially) overlapping windows across the space of the relevant collective variables, in order to limit the scale of the free energy features that the system needs to thermally explore. A full free energy surface is then reconstructed by stitching together samples from the individual windows. While successful, this approach requires system-specific definition of the thermodynamic windows and laborious post processing, making it challenging for non-experts.\\

Alternatively, a single biasing potential can be designed to globally counteract the free energy profile. However, that ideal bias is unknown at the outset; it must be initially set using intuition, and then iteratively refined in subsequent simulations. The fast-iteration limit of refinement is on-the-fly update, where an optimal bias is progressively learned in a single simulation rather than optimized through separate runs -- this is the idea behind metadynamics (MetaD)\cite{Laio2002,Bussi2020}.\\

In MetaD,  a bias is constructed from the history of observed configurations, which discourages revisiting of previously sampled states. This process encourages iteratively wider exploration of the state space, eventually enabling transitions over the free energy barriers separating local minima. Even for systems with a single free energy minimum, MetaD enables sampling of high free energy states, an ability that would be particularly useful to probe force-response in DNA nanomachines and mechanical sensors \cite{Zhou2014,Zhou2015,Hudoba2017,Dutta2018}. MetaD simulations can benefit from GPU acceleration and a natural parallelisation route through multi-walker metadynamics\cite{Raiteri2006}. The well-tempered variant of MetaD\cite{Barducci2008} limits the maximum correction to the free energy landscape, preventing irreversible disassembly.  Finally, there is no need to run simulations with multiple thermodynamic windows, as in umbrella sampling, simplifying execution and post-processing, and eliminating some concerns about hysteresis \cite{Zhu2012}.\\

Here we introduce an implementation of well-tempered MetaD in the oxDNA simulation framework, which offers a viable route for the rapid assessment of conformation free energy landscapes in DNA nanotechnology. To demonstrate the validity of the technique we applied it to four case-studies where conventional MD would be unable to probe the relevant landscapes: (\emph{i}) the compression-induced buckling in duplex DNA, (\emph{ii}) conformer transitions in bistable Holliday junctions \cite{Lilley2000}  and (\emph{iii}) switchable tiles\cite{Song2017}, and (\emph{iv}) force response in an origami compliant joint, where conformation is prescribed by balancing competing forces\cite{Zhou2014}. For systems (\emph{ii}) and (\emph{iv}) we compared simulation outcomes with experimental observations, finding quantitative agreement. Overall, we demonstrated that MetaD, as applied to oxDNA, can effectively sample transitions between multistable systems and facilitate the computational characterization of highly deformable designs, all in an automated fashion that requires limited system-specific user input. This tool could therefore be highly valuable in computer-assisted design and assessment pipelines for reconfigurable DNA nanostructures.\\

\section{Results and discussion}

{
\subsection{Principles of metadynamics}

Here, we give a brief overview of the principles and implementation of MetaD. A complete theoretical description can be found in Bussi \emph{et al.} \cite{Bussi2020}. The objective of MetaD is to map a free energy landscape from molecular simulation. As landscapes typically have high dimensionality, for human interpretation the free energy is projected onto a set of lower-dimensional coordinates or \emph{collective variables} ($\vec{s}$), defined as functions of the coordinates of the simulated system ($\vec{q}$).\\

In principle, long trajectories sampled from Monte Carlo (MC) or MD, can be used to infer free energy landscapes from state-occupancy histograms. The projection of the free energy onto {a discretized coordinate $\vec{s}_0$} can then be estimated as ${\Delta G(\vec{s}_0) \simeq -k_BT\log N({\vec{s}_0})+c}$, where $c$ is an immaterial constant, $k_B$ is the Boltzmann constant, $T$ is the temperature, and $N({\vec{s}_0})$ is the number of samples in the histogram bin centred at $\vec{s}_0$ \cite{Frenkel2002}. However, convergence of this unbiased approach is practically unfeasible for many macromolecular and DNA nano-systems owing to the presence of thermally inaccessible configurations, that frequently separate multiple metastable minima.\\

MetaD  generates a history-dependent bias that progressively flattens the free energy landscape, thus rendering high free energy regions accessible, and enabling efficient sampling.\\ 

A MetaD simulation proceeds as follows. The system is initialized and simulated (with either MD or MC algorithms) using a potential defined as $U_t(\vec{q})=U(\vec{q})+B_t(\vec{s}(\vec{q}))$, where $U(\vec{q})$ is the unbiased potential and  $B_t(\vec{s}(\vec{q}))$ the time-dependent bias. The index $t$ indicates the number of MetaD \emph{iterations} performed, each iteration consisting of $\tau$ (MD/MC) time-steps. The bias is initialized as $B_{t=0}(\vec{s}(\vec{q}))=0$, and updated after each iteration to counteract the projection of the free energy onto $\vec{s}$. To calculate the updated bias, the instantaneous value of $\vec{s}$ is evaluated, termed $\vec{s}_t$. The bias is then updated through the addition of a Gaussian potential centred at $\vec{s}_t$, which discourages the system from revisiting its current state

\begin{align}\label{eq_bias}
    {B_{t+1}(\vec{s})=B_{t}(\vec{s})+w\exp\bigg({-\frac{(\vec{s}_t - \vec{s})^2}{2\sigma^2}}\bigg)}.
\end{align} 

In equation~\ref{eq_bias}, $\sigma$ is the width of the deposited Gaussian, while the parameter $w$ controls the rate at which the free energy wells are filled. In the earliest version of MetaD, also known as direct MetaD, $w$ was set to a constant value~\cite{Laio2002}, resulting in a $B_t$ which oscillates rather than converging \cite{Bussi2020}. Alternatively, convergence of $B_t$ can be guaranteed by reducing $w$ in areas that are already strongly biased, an approach known as well-tempered metadynamics \cite{Barducci2008}, which we adopt throughout this work. In well-tempered MetaD, the time dependent amplitude of the Gaussian, $w_t$ is given by

\begin{align} \label{eq_wt}
w_t=A\exp\bigg(-\frac{B_t({\vec{s}_t})}{k_B\Delta T}\bigg).
\end{align}

In equation~\ref{eq_wt}, $\Delta T$ is an additional hyperparameter with units of temperature, which controls the strength of tempering. High values of $\Delta T$ correspond to weak tempering, where forces are allowed to accumulate, with $\Delta T \rightarrow \infty$ approaching conventional MetaD (constant $w$). Conversely, low values of $\Delta T$ correspond to systems which quickly taper their bias, with the $\Delta T\rightarrow0$ limit corresponding to unbiased sampling. The value of $A$ controls the initial bias-height increment. $\Delta T$ and $A$ are set at the start of the simulation, alongside the other parameters ($\sigma$ and $\tau$) and the collective variables. With well-tempererd MetaD, at long times, the value of $B_t(\vec{s})$ provably converges to a fraction of the projection of the free energy onto the collective variable (up to an immaterial constant, $c$)\cite{Dama2014b}
\begin{align}\label{eq_conv}
\lim_{t\to\infty} B_t(\vec{s})= -\frac{\Delta T}{\Delta T+T}\Delta G(\vec{s}) + c.
\end{align}
An estimate of $\Delta G(\vec{s})$ can be therefore be acquired from the converged bias\cite{Bonomi2009}.
Additionally, this equation illustrates the physical interpretation of $\Delta T$. After convergence, the residual (\emph{i.e.} uncorrected) free energy felt by the system is ${B_t + \Delta G = \frac{T}{T+\Delta T}\Delta G} $, implying that $T+\Delta T$ can be interpreted as the effective temperature experienced along a collective variable \cite{Bussi2020}.\\
While equation \ref{eq_conv} enables estimation of $\Delta G$, a preferred route is that of directly extracting the sought free energy from configuration histograms of simulation runs biased with the asymptotic $B_t$. This approach will be used to derive free energy landscapes in the remainder of this article, unless specified otherwise.\\

Supplementary Note 1 and figure S1 demonstrate the implementation of MetaD to a basic one-dimensional example, while, in the reminder of this paper, we illustrate its applications to mapping deformation free energy landscapes for increasingly complex DNA nano-systems, simulated with MD and  the coarse grained oxDNA force field.\\

Information on the implementation of MetaD in oxDNA, and specific simulation details for all case studies can be found in the Methods section and tables S1 and S2.\\

{ While convergence of (well-tempered) MetaD is very robust, the free parameters $\sigma$, $A$ and $\tau$, alongside system-dependent features such as physical size, intrinsic diffusion times and collective-variable dimensionality, have been shown to influence errors in free energy estimates and convergence timescales~\cite{Laio2005,Bussi2006}. In the Methods we discuss these factors and other practical considerations that guided our parameter choice.}\\
}

\subsection{Bending and buckling free energy of a DNA duplex}
{
In this section we demonstrate the application of MetaD to coarse-grained oxDNA simulations using a simple case study:  the response of double-stranded (ds)DNA under strong bending. A similarly simple application is discussed in ref.~\citen{Sicard2015}, which explores bubble formation in a basic bead-and-spring model of a DNA duplex.}\\

dsDNA is often thought of as a Worm-like Chain (WLC) -- an elastic beam whose bending energy is quadratic in local curvature, much like a macroscopic beam. If the ends of such a duplex are compressed together, then the WLC model predicts that the curvature will increase everywhere. However, experimental evidence indicates that under a sufficient compressive load, a short dsDNA duplex will not bend continuously. Instead it will buckle, and in this buckled state there will be a single point of high curvature -- a kink~\cite{Vologodskii2013}. Experimental observations of force induced kinking have been identified for a DNA-based molecular vice in fluorimetry experiments \cite{Fields2013}, in the vulnerability of dsDNA minicircles to single-stranded (ss)DNA-specific enzymatic degradation  \cite{Du2008}, and also \emph{via} AFM of said minicircles \cite{Pyne2021}. Similarly, kink formation under conditions of end-to-end compression has also been observed in atomistic simulation \cite{Lankas2006}, and with the oxDNA force field \cite{Harrison2015,Harrison2015a}. Both atomistic and coarse-grained simulations indicate that the origin of kinking is a local break in the continuity of coaxial stacking in the helix \cite{Harrison2015,Harrison2015a,Lankas2006}, and may be also associated with the loss of a Watson-Crick bond. Here we have used sampling of DNA kinking as simple test application of MetaD in oxDNA.\\

\begin{figure*}[ht!] 
\centering    
\includegraphics[width=\textwidth]{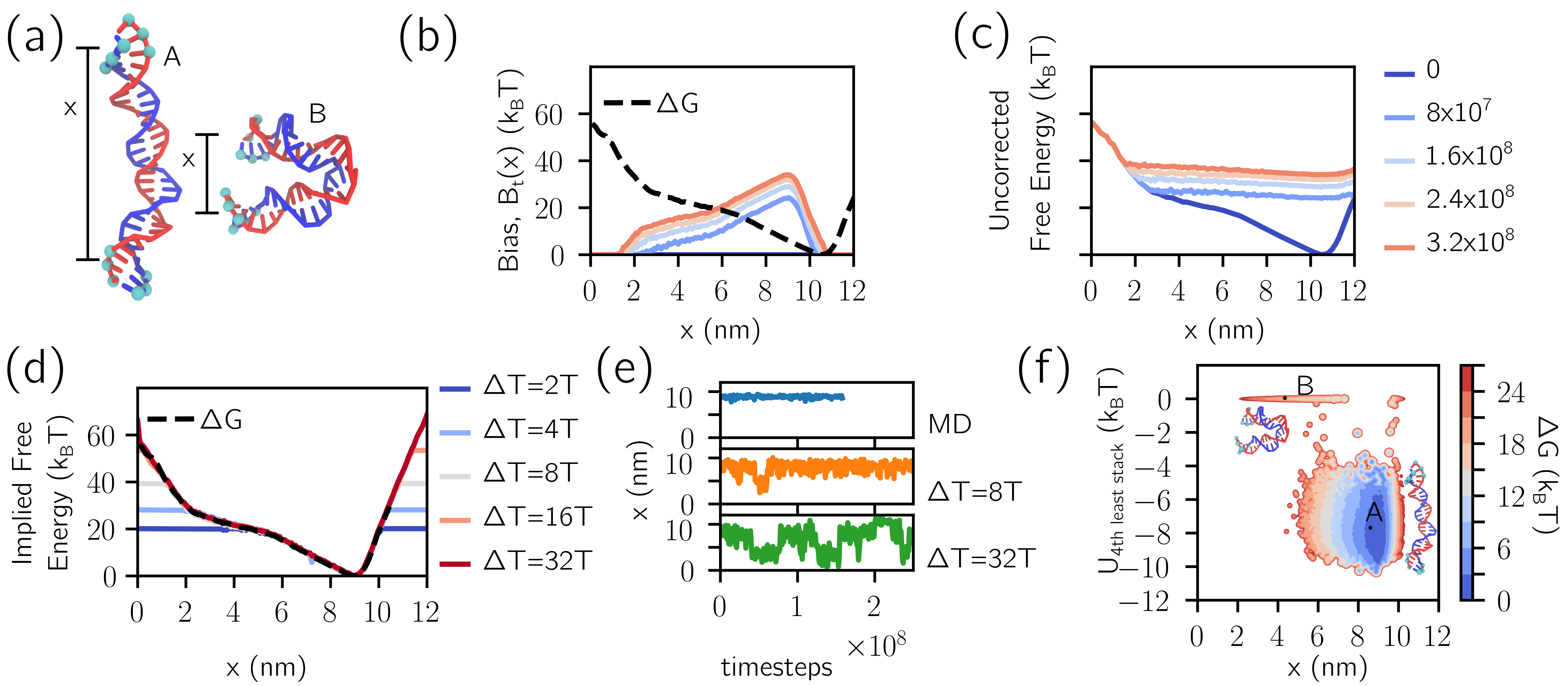}
\caption{MetaD enables automated sampling of dsDNA buckling. \textbf{(a)},~Snapshots of unbuckled {(left, A)}, and buckled {(right, B)} configurations of a DNA duplex from MetaD simulation. The buckled state features disrupted stacking roughly in the center of the duplex. The distance $x$ between the centers of mass of the two collections of six cyan beads was used as the collective variable. \textbf{(b)},~The time dependence of the bias $B_t$ for a system with $\Delta T={8T}$. Also plotted is $\Delta G$, the unbiased potential experienced by the system ({black} dashed line). $B_t$ is initially flat, and the it builds up according the history of visited configurations {(equation~\ref{eq_bias})}. \textbf{(c)},~The simulation experiences a potential equal to $\Delta G + B_t$ -- the uncorrected potential. As illustrated, the initial uncorrected potential is sharply varying, but then it progressively flattens as the bias grows, enabling  access to a wider $x$-range. { Different colours mark different numbers of MD time-steps, as indicated in the legend which applies to both panels \textbf{b} and \textbf{c}}. \textbf{(d)}, Implied free energy from the asymptotic $B_t$, for varying $\Delta T$ {(equation~\ref{eq_conv})}. $\Delta G(x)$ is plotted as a black dashed line. \textbf{(e)},~Trajectories of the collective variable $x$ for $\Delta T=0$ (ordinary MD), $\Delta T={8T}$, and $\Delta T={32T}$.  \textbf{(f)},~Two-dimensional free-energy landscape acquired from biased MD simulation. The $y$ axis indicates $U_{\text{4th least stack}}$, which rises to $0$ only if at least four non-terminal nucleotides lack stacks -- \emph{i.e.} a buckled state. Locations marked as A and B correspond to the snapshots in \textbf{a}.}
\label{fig:meta.dsDNA}
\end{figure*}


{We wish to apply MetaD to calculate how free energy varies with the end-to-end distance of a short duplex DNA, which, due to the complex buckling transition, is impossible to calculate analytically.} Figure~\ref{fig:meta.dsDNA}\textbf{a} shows snapshots of the unbuckled (left, A), and buckled (right, B) configurations of the duplex. The distance $x$ between the centers of mass of the two collections of six cyan beads was used as a collective variable { onto which the free energy is projected and the MetaD bias $B_t$ applied}. In figure~\ref{fig:meta.dsDNA}\textbf{b}, the time evolution of $B_t(x)$ has been plotted, along with reference free energy $\Delta G(x)$ -- the true energetic cost to bend the duplex. 
{ The uncorrected potential $B_t(x)+\Delta G(x)$, \emph{i.e.} the residual potential felt by the system, progressively flattens as $B_t(x)$ evolves according to equations~\ref{eq_bias} and~\ref{eq_wt} to counteract $\Delta G(x)$ (figure~\ref{fig:meta.dsDNA}\textbf{c}).}\\ 

In the examples given in figure~\ref{fig:meta.dsDNA}\textbf{b} and \textbf{c}, the tempering parameter $\Delta T$ {(equation \ref{eq_wt})} has been set to $8T$, so that the bias converges to $-\frac{8}{9}\Delta G$  {(equation \ref{eq_conv})}; \emph{i.e.} the asymptotic uncorrected potential is $\frac{1}{9}$ of the true value. Figure~\ref{fig:meta.dsDNA}\textbf{d} shows free energy implied according to equation \ref{eq_conv} for different values of $\Delta T$, compared with the reference free energy (see figure S2 for proof of convergence of the biases). As expected, larger $\Delta T$ values produce accurate estimates of $\Delta G$ away from the minimum. \\

In figure \ref{fig:meta.dsDNA}\textbf{e}, the time varying values of $x$ are given for three different values of $\Delta T$. Under conventional MD ($\Delta T=0$, blue), only the unbuckled state is sampled. When metadynamics is turned on ($\Delta T=8T$ or $32T$, yellow and green respectively), the bias repels the system from previously visited configurations, resulting in a wider exploration in the unbuckled free energy minimum. From {$\approx 5\times10^7$ MD} time-steps, both biased systems begin exploring the buckled state at smaller $x$-values, only briefly for $\Delta T=8T$ and more persistently for $\Delta T=32T$. The latter simulation then experiences frequent transitions between buckled and un-buckled states.\\ 


In the one-dimensional free energy profile projected along $x$, the configuration corresponding to the buckled and un-bucked states does not appear separated by a free-energy barrier. However, such a potential barrier exists, and can be visualized along alternative coordinates, as shown with the two-dimensional free-energy landscape in figure~\ref{fig:meta.dsDNA}\textbf{f}. Here, we introduce a second collective variable, $U_\text{4th least stack}$, defined as the value of the fourth weakest stacking interaction, which we expect to increase as the duplex buckles and a kink forms. Indeed, in figure~\ref{fig:meta.dsDNA}\textbf{f} we observe two distinct states: a broad minimum at large $x$ and finite (negative) $U_\text{4th least stack}$, associated with the un-buckled duplex, and a second { minimum} centred at smaller $x$ and with $U_\text{4th least stack}=0$, corresponding to the buckled duplex.  Transitions between the two minima are not effortless even with $\Delta T=32T$, but good sampling is possible with many replicas which are run simultaneously, sharing and contributing to the same bias {(see Methods)}.\\

$\Delta T$ can be used to control which parts of the free energy landscape should be explored, and the trade-off between sampling a large region of collective variable sparsely, or a small region well. It can also be used to eliminate sampling of states which may be undesirable. For example, a low value of $\Delta T$ could be use to prevent sampling of kink formation if the objective were to identify only bending close to the free energy minimum.\\

\subsection{Two-dimensional isomerization landscape of bistable motifs}

\begin{figure*}[ht!] 
\centering    
\includegraphics[width=\textwidth]{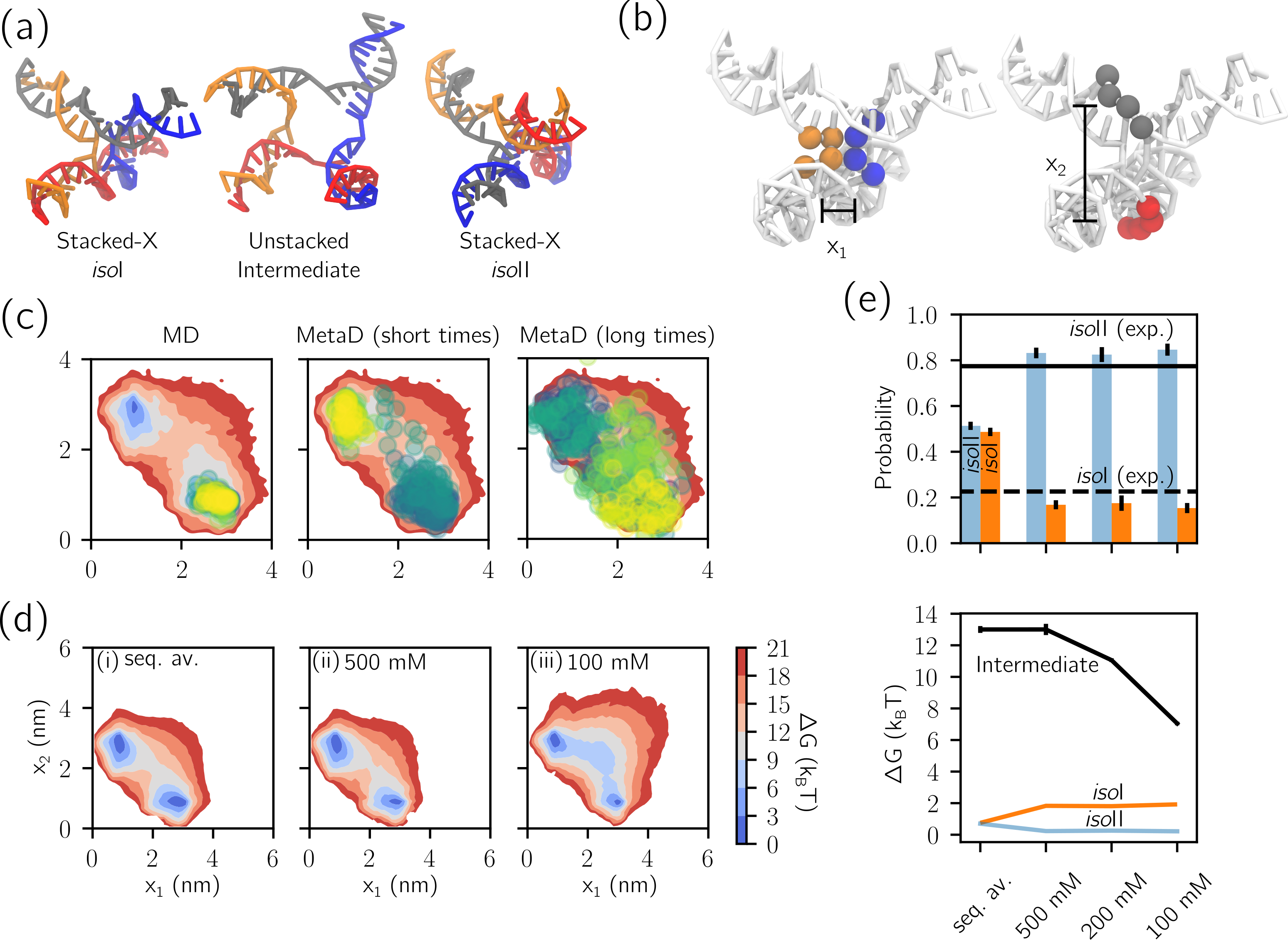}
\caption{MetaD enables sampling of the isomerization free energy landscape of bistable Holliday junctions. \textbf{(a)},~A Holliday junction consists of two quasicontinuous duplexes joined by a crossover, as illustrated in the snapshot. There are two dominant conformers, one where the grey and red strands are fully stacked (left, \emph{iso}I), and another where the orange and blue strands are fully stacked (right, \emph{iso}II). An unstacked structure is believed to be the intermediate (center). \textbf{(b)},~For MetaD simulations we used a two-dimensional collective variable, ($x_1$, $x_2$), where $x_1$ is the distance between the centers of mass of the orange and blue sets of beads, while $x_2$ is the distance between centers of mass of the red and grey beads. \textbf{(c)},~A 5$\times 10^7$ time-step trajectory simulated under unbiased MD (left), and one of the same duration collected with MetaD (center), both overlaid with the ($x_1$, $x_2$) free energy profile. Simulating over 10 times this period in MetaD results in many transitions between conformers, enabling accurate sampling of the free energy landscape (right). Dots of different colors indicate sampled configurations.
\textbf{(d)},~Free energy surfaces corresponding to (i)  the sequence-averaged model at 500 mM ionic strength, and the sequence-specific model~\cite{Sulc2012} at (ii) 500 mM NaCl and (iii) 100 mM NaCl. See figure S3 for data on the sequence-specific model at 200 mM ionic strenght. In each case, there are two minima, corresponding to the two stacked-X conformers,  and a saddle-point region associated to the intermediate. \textbf{(e)}, (Top) Probabilities for the \emph{iso}I and \emph{iso}II states for the four studied systems (bars), compared with experimental values ({ dashed (\emph{iso}I) and solid (\emph{iso}II)} black lines)\cite{Joo2004}. (Bottom) Free energies of the \emph{iso}I, \emph{iso}II {(color-coded as in panel \textbf{d})} and intermediate states {(black)}. The free energy of the intermediate falls by approximately 7 $k_BT$ between the systems with 500 mM and 100 mM ionic strength, consistently the experimentally observed phenomenon of faster isomerization at low salt concentrations. Error bars are the standard error based of 6 replicas {(too small to see for \emph{iso}I and \emph{iso}II)}.
}
\label{fig:meta.nanostar}
\end{figure*}

While in our first case study a single collective variable was sufficient to bias the simulation and extract the sought information, it is often the case for (relatively) more complex DNA architectures that multi-dimentional free energy landscapes need to be explored. To this end, Holliday junction isomerization provides a useful case study. The immobile Holliday junction was the first non-trivial DNA motif to be intentionally constructed \cite{Kallenbach1983}, and consists of four helices joined at a central four-way junction. Its configuration in the presence of divalent, or high concentrations of monovalent cations is that of two quasi-continuous helices joined at a strand crossover location. This is referred to as the stacked-X configuration \cite{Lilley2000}, and is shown in figure~\ref{fig:meta.nanostar}\textbf{a} (left, right). In the absence of such cations, the construct acquires an unstacked planar configuration, where each of the four arms can move flexibly about the central junction (figure~\ref{fig:meta.nanostar}\textbf{a}, center)\cite{Lilley2000}.\\

Stacked-X Holliday junctions can exist in two conformers,  distinguished based on which of the four helices are stacked at the junction (figure \ref{fig:meta.nanostar}\textbf{a}, left and right). These conformers, previously referred to as \emph{iso}I and \emph{iso}II\cite{Lilley2000}, are structurally equivalent if base-sequence is ignored, while asymmetry of base pairs at the junction results in one conformer being favoured. In the presence of MgCl$_2$, each conformer is long lived -- single molecule Förster Resonance Energy Transfer (FRET) experiments indicate lifetimes of milliseconds to seconds \cite{Joo2004}. Consequently, sampling transitions between the two conformers is intractable with typical molecular simulation approaches. For alternative representations of DNA, transition sampling has required running the simulations at vastly increased temperature \cite{Yu2004}, or using a coarse-grained force field which overestimates the stability of the transition state \cite{Wang2016}. The properties of the oxDNA representation of a Holliday junction have been explored previously\cite{Snodin2019}. However, transitions, and sequence dependent conformer probability have remained unexplored due to the non-ergodicity of this system under conventional MD sampling. Here we show that, using a two-dimensional reaction coordinate,  MetaD can successfully sample conformer transitions and determine the relative conformer stability. \\

The particular structure investigated here is similar to the J3 junction, previously  characterized experimentally~\cite{Joo2004}, with the only difference being the dsDNA ``arms'' have been truncated to 11 bp to enable faster simulations (see table S3 for sequences). 
To favour the formation of an unstacked intermediate state, thus enabling transitions between conformers, we used a two dimensional collective variable, corresponding to the two diagonal distances across the Holliday junction ($x_1$ and $x_2$ in figure~\ref{fig:meta.nanostar}\textbf{b}). In the stacked-X state, one of these distances takes a high value, corresponding to the width of the junction, while the other takes a low value, corresponding approximately to the axial rise of two base pairs. Meanwhile, the planar transition state corresponds to high values of both collective variables.\\

To demonstrate the enhanced sampling made possible in MetaD versus conventional MD, in figure~\ref{fig:meta.nanostar}\textbf{c} we have plotted small sections of trajectories for both techniques. While a trajectory simulated under MD remains stuck in a single minimum (figure~\ref{fig:meta.nanostar}\textbf{c}, left), using MetaD it is able to escape and sample several transitions (figure~\ref{fig:meta.nanostar}\textbf{c}, center and right).\\

Free energies projected onto the collective variables are plotted in figure~\ref{fig:meta.nanostar}\textbf{d}, as acquired from MD using an asymptotic bias from MetaD. Illustrations of the similarity between the converged MetaD bias and the free energy from biased MD simulation are given in figure S3. Two different constructs were tested, one which ignores base identity by using sequence-averaged parameters and one utilizing the sequence-dependent force field~\cite{Sulc2012}. The former construct was simulated at 500 mM ionic strength, while the latter at three different ionic strengths (500, 200 and 100 mM). As expected, while the sequence-averaged calculations produce a symmetric landscape, sequence dependence results in asymmetry, with one conformer being favoured over the other.  Additionally, we observe that  the intermediate region between the two conformer minima flattens at lower ionic strengths. This region corresponds to the unstacked intermediate, which thus appears to be favored by a reduction in salt concentration.\\

Experiments indicate that in conditions of 50~mM MgCl$_2$ (ionic strength 150 mM), the J3 junction will display the \emph{iso}II conformer 77.4\% of the time\cite{Joo2004}. In figure \ref{fig:meta.nanostar}\textbf{e} we show simulated conformer probability for the four studied systems. For the sequence-specific model we find that conformer probability is independent on ionic strenght, and in quantitative agreement with experimental observations. This agreement with experimental results is intriguing, as stacking interactions in oxDNA have not been parameterized to reproduce Holliday junction conformer prevalence, but instead the melting transitions of duplexes and hairpins based on the Santa-Lucia parameters \cite{Snodin2015,santalucia1998}. The reproduction of conformer probability is further validation of the oxDNA model of stacking. Definitions of the stacked and transition states are discussed in the Methods.\\

Figure \ref{fig:meta.nanostar}\textbf{e} (bottom) shows how the free energies of the two conformers, as well as the unstacked intermediate state, depend on ionic strength. While the values for the stacked-X configurations remain constant, ion concentration is critical in controlling the free energy of the unstacked intermediate, as previously noted. Indeed, as ionic strength falls { from 500 to 100\,mM}, the relative free energy of the intermediate decreases by $\approx$ 7$k_BT$, making it  approximately three orders of magnitude more likely. This effect is due to increased electrostatic repulsion associated with a less concentrated electrolyte; the stacked Holliday junction has a high density of negative charge, and is therefore disfavored when electrostatic screening is reduced.\\ 

An association can be made between the stabilization of the intermediate at lower ionic strengths and the increase in conformer interconversion rate, defined as sum of the rates of \emph{iso}I$\rightarrow$\emph{iso}II and \emph{iso}II$\rightarrow$\emph{iso}I \cite{Joo2004} transitions. The latter, as determined experimentally for a junction of slightly different sequence, rises from 20\,s$^{-1}$ at 2\,M Na$^+$ to 800\,s$^{-1}$ at 400\,mM Na$^+$. A similar increase is observed in systems with magnesium counterions if their concentration is dropped from 100\,mM (interconversion rate 10 s$^{-1}$) to 7\,mM (interconversion rate 500 s$^{-1}$). By assuming direct proportionality between the interconversion rate and probability of the unstacked intermediate, oxDNA would predict that reducing ionic strenght from 500\,mM to 200\,mM would result in a 7-fold increase in isomerization rate, while reducing the ionic strenght further, to 100\,mM, would accelerate isomerization by a factor of 1000. However,  it should be noted that these considerations are purely qualitative, and rare-event sampling techniques which do not create fictitious dynamics \cite{Allen2009} are typically required to make definite claims about transition rates and paths. \\

As an additional example, in Supplementary Note 2 and figures S4-S6 we test MetaD on a second bistable unit where transition between two conformers requires breaking of stacking interactions. This tile has been utilized as elementary unit of re-configurable origami that can spatially relay information through the propagation of conformational transitions along an array of units\cite{Song2017}. Similar to the case of the Holliday junction, we are able to efficiently reconstruct the transition free-energy landscape utilizing both a one- and a two-dimensional collective variables for biasing, which would be not be viable with unbiased MD. We are also able to gather information on the transition pathway between conformers; however, this needs to be interpreted with care owing to potential artefacts introduced by the biasing potential.\\

\subsection{Bending free energy of a compliant origami joint}

\begin{figure*}[ht!] 
\centering    
\includegraphics[width=\textwidth]{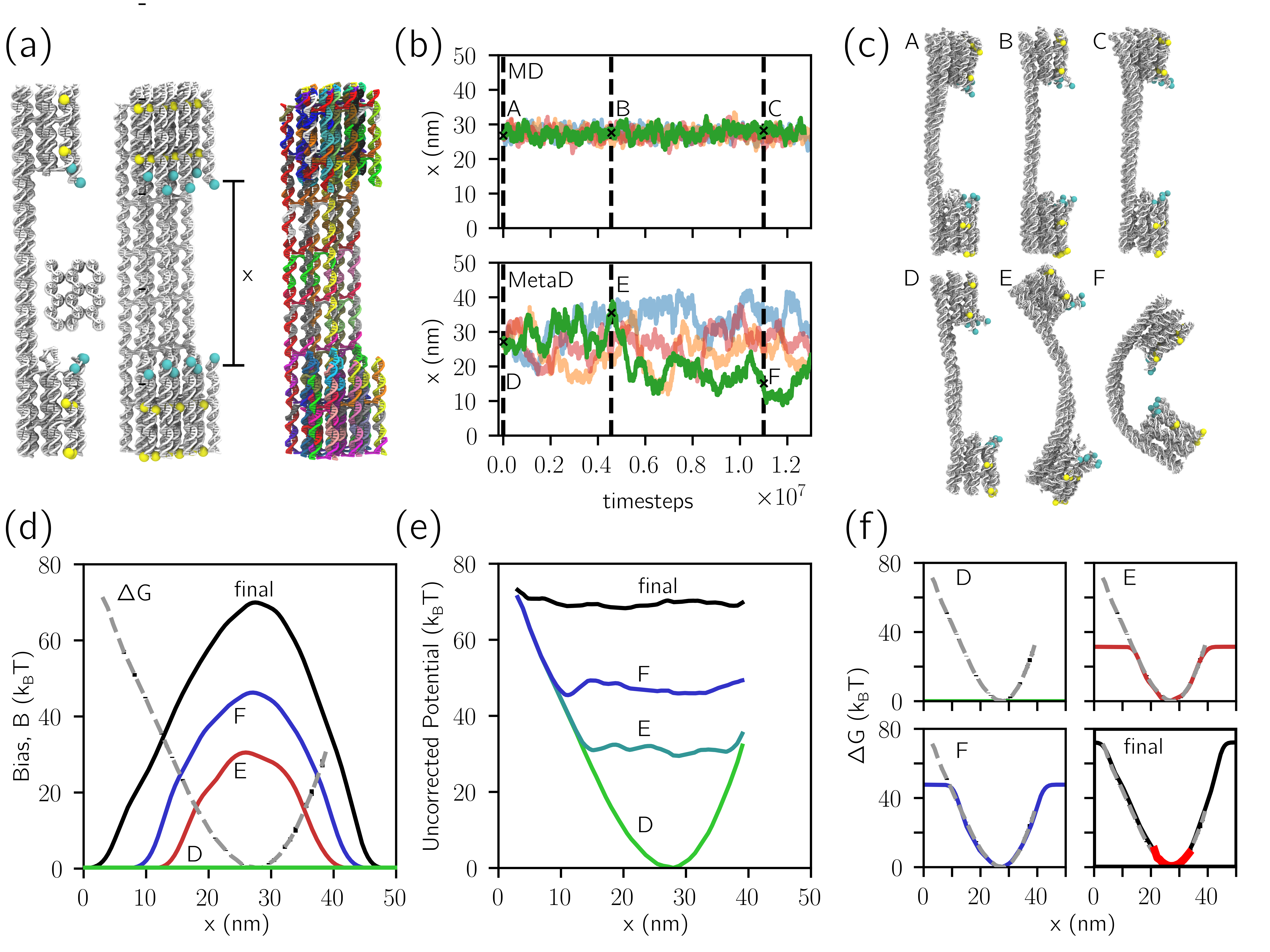}
\caption[Metadynamics enables sampling of mechanically stressed states]{Metadynamics enables sampling of mechanically stressed states in DNA origami. \textbf{(a)},~A mechanically compliant DNA origami joint, truncated here from its experimental realization~\cite{Zhou2014}. The cross section of the 18 helix bundle is also shown. The yellow beads were used as references in the bending angle $\phi$ (figure~\ref{fig:meta.origami2}). The collective variable $x$ is defined from the distance between the centers of mass of the top collection of six cyan particles, and the six at the bottom. { Individual staples and scaffold, whose routing is depicted in figure S7, are color coded in the right-hand-side image.} \textbf{(b)},~MD simulation (top) only samples around the free energy minimum of the compliant joint, yielding little information about the force required to actuate it. MetaD simulation (bottom) learns to sample a wider range of configurations. The illustrated trajectories correspond to approximately half the total time sampled in MetaD simulations. Different colors indicate parallel replicas that for MetaD contribute to, and experience, the same bias potential. \textbf{(c)},~Snapshots are illustrated from MD simulation (top), and MetaD simulation (bottom). \textbf{(d)},~Time evolution of the MetaD bias $B_t$ {(solid lines)}. Letters refer to biases at simulation times corresponding to those illustrated in \textbf{b}. The long time limit bias {(final)} is also shown. The true free energy $\Delta G$ is shown as a dashed grey line { with black $1\sigma$ errorbars (often too small to see)}. \textbf{(e)}, The sum of $\Delta G$ and $B_t$ -- the uncorrected potential -- is plotted for different simulation times. \textbf{(f)}, Free energy profiles (continuous lines) as implied from $B_t$ are plotted alongside $\Delta G$ ({gray} dashed, 1\,$\sigma$ error bars), demonstrating convergence. { The thicker red curve in the bottom-right sub-panel represents the free energy profile as determined from direct sampling of unbiased trajectories in \textbf{b} (top).}}
\label{fig:meta.origami1}
\end{figure*}

The principle of compliant mechanism design is to control mobility and mechanical properties \emph{via} local thinning of material, rather than through rigid body linkages \cite{Howell2013}. This is a popular approach when designing DNA origami with an intended pattern of motion, where the number of helices is reduced in regions of the structure where compliance is desired \cite{Zhou2014,Zhou2015}. Here, we consider a DNA origami compliant joint as a useful case study for the mechanical predictions of the MetaD approach. The joint has been previously characterized experimentally \cite{Zhou2014}, and computationally with oxDNA using unbiased MD\cite{Shi2017}. The latter study demonstrates that oxDNA can accurately capture the shape of compliant DNA structures, although it under-predicts the width of conformational distributions\cite{Shi2017}. We simulated a truncated version of the experimentally realized joint, illustrated in figure~\ref{fig:meta.origami1}\textbf{a}, where truncation improves computational efficiency. The joint is composed of two 18-helix bundles, connected by a thinner 6-helix layer. Consequently, bending will preferentially occur in plane, localized to the thinned layer. See figure S7 for the caDNAno routing of the device.\\

Through MetaD simulations, we can explore the bending free energy of the joint, sampling highly deformed configurations inaccessible to conventional MD. We bias the simulations using the collective variable $x$, defined as the average distance between the centers of mass of top and bottom collections of cyan beads, illustrated in figure~\ref{fig:meta.origami1}\textbf{a}. As demonstrated in figure~\ref{fig:meta.origami1}\textbf{b} (top), MD explores states only close to the free energy minimum, physically corresponding to an unstressed six-helix section. Snapshots corresponding to these trajectories are illustrated in figure~\ref{fig:meta.origami1}\textbf{c} (top). By contrast, MetaD initially explores the free-energy minimum, and then is pushed by the bias to explore other regions of configuration space (figure~\ref{fig:meta.origami1}\textbf{b}, bottom). Trajectories in MetaD widen with time, not just because of diffusion, but because the free energy landscape felt by the system is progressively flattened. Snapshots illustrated in figure~\ref{fig:meta.origami1}\textbf{c} (bottom) indicate the sampling of high free energy states which would never have been reached in unbiased MD.\\


\begin{figure*}[ht!] 
\centering    
\includegraphics[width=\textwidth]{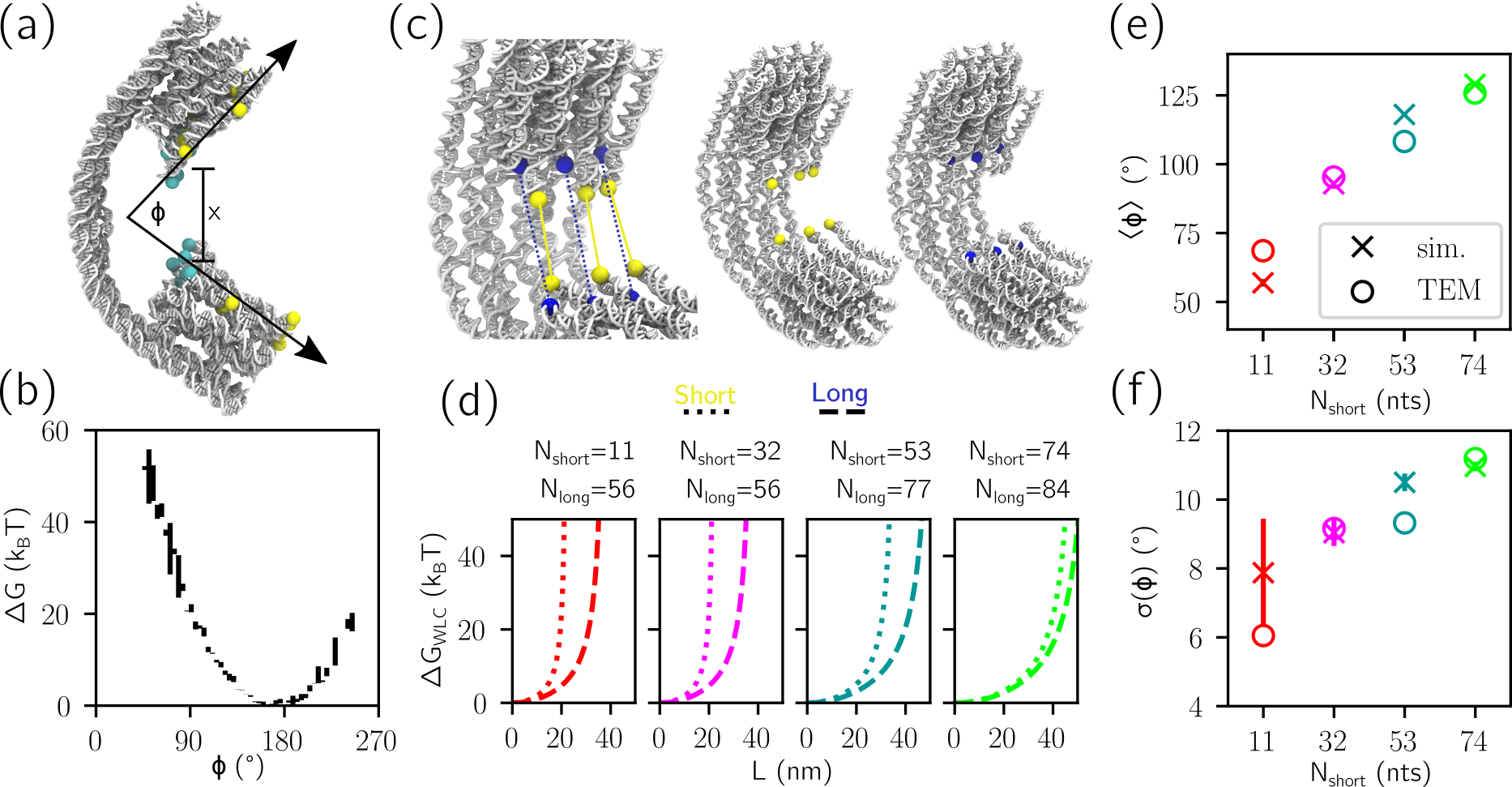}
\caption[Metadynamics allows prediction of the mechanical response to an external force.]{Metadynamics allows prediction of the mechanical response of a DNA origami to an external force. \textbf{(a)},~Definition of the angle $\phi$. Straight lines defining the angle are those passing through the centers of mass of the groups of yellow beads nearest and furthest from the joint. See further details in the Methods.  \textbf{(b)},~The bending free energy profile against $\phi$ { as estimated with MD using the converged MetaD bias}.  \textbf{(c)}, Renders of the ssDNA connections across the joint, as implemented experimentally~\cite{Zhou2014}. There are three short ssDNA sections (yellow), and three long sections (blue). \textbf{(d)},~Free energy profiles of WLCs for each of the ssDNA distributions under study.  \textbf{(e)}, Predictions of mean compliant joint angle, $\langle\phi\rangle$, compared to those from Transmission Electron Microscopy (TEM) experiments~\cite{Zhou2014}. \textbf{(f)}, Predictions of the standard deviation of angle width, $\sigma(\phi)$, compared to those from TEM. { Data points are color-coded as for the corresponding WLC free energy curves in panel \textbf{b}}. All error bars represent the standard errors evaluated from different replicas, as discussed in the Methods; those in \textbf{e} are smaller than the symbols.}
\label{fig:meta.origami2}
\end{figure*}

The time dependency of the bias is illustrated in figure~\ref{fig:meta.origami1}\textbf{d}, with letters corresponding to the times marked in figure~\ref{fig:meta.origami1}\textbf{b}. Notice, the structural similarity of the reference free energy to the final bias reached by the simulation. Similarly, figure~\ref{fig:meta.origami1}\textbf{e} shows the time-evolution of the uncorrected potential, 
 ${B_t(x)+\Delta G(x)}$, which progressively flattens as previously noted in figure \ref{fig:meta.dsDNA}, while figure~\ref{fig:meta.origami1}\textbf{f} shows how the implied potential converges to the reference curve. { In figure~\ref{fig:meta.origami1}\textbf{f} (bottom right) we also show the free energy as determined from direct sampling of un-biased MD simulations (panel \textbf{b}, top), which expectedly are only able to reconstruct the profile for thermally accessible configurations.} \\

Experimental investigations of the compliant joint have relied on a bending angle, rather than a distance, to classify the deformation state of the nanomachines. For direct comparison, and thus to demonstrate the predictive power of the oxDNA MetaD approach, we have defined the bending angle $\phi$ as illustrated in figure~\ref{fig:meta.origami2}\textbf{a}, closely matching  the definition used in Transmission Electron Microscopy (TEM) experiments \cite{Zhou2014}. The associated bending free energy profile is plotted in figure~\ref{fig:meta.origami2}\textbf{b} -- note once more how MetaD enables sampling high free energy states associated to extreme bending, with free energies reaching $\sim 60\,k _BT$ above the ground state.\\

In experiments, controlled bending of the joint has been induced through the addition of ssDNA segments, bridging the 18-helix bundles across the flexible section of the joint at the locations illustrated in figure~\ref{fig:meta.origami2}\textbf{c}. Three strands join yellow beads (each containing $N_\text{short}$ nucleotides), and another three, with possibly different lengths, join blue beads (each with $N_\text{long}$ nucleotides). The segments act as entropic springs, bending the 6-helix bundle and determining the configuration (and flexibility) of the joint. The bending state can thus be controlled by changing the length and number of the springs\cite{Zhou2014}.\\

Besides assessing the flexibility of the unconfined joint, a useful role for simulations would be that of predicting the mean bending angle that results from a given set of ssDNA springs, so to inform experimental design. To this end, one approach would be to perform separate simulations for many possible lengths of ssDNA springs \cite{Shi2017}, and then manufacture the system whose behavior is closest to the desired outcome. While this is computationally costly, it is the only possible approach when states far from the location of minimum free energy cannot be sampled.\\
MetaD, instead, unlocks a much more efficient approach thanks to its ability to sample with a single simulation the entire distribution of angles, as we have shown.  Once this free energy profile is known in the absence of any ssDNA, one can indeed analytically account for the constraints imposed by ssDNA springs. Specifically we can predict the bending angle distribution, by reweighting the distributions from biased MD to account for the energetic contribution of the springs, as described in the Methods. Each ssDNA section is modelled as a separate WLC between attachment points, and free energies from each contribute to the reweighting. Figure \ref{fig:meta.origami2}\textbf{d} illustrates the free energy contribution for each combination of long and short chains used here, as a function of extension. This strategy offers an efficient alternative to determine the bending-angle distribution of the joint for any choice of ssDNA springs, ensuring that the inverse problem of designing ssDNA sections to produce a given angle is approachable. Similarly, it offers a way to estimate the flexibility of the joint under applied force, useful if it were later used in a load bearing application.\\

Figure~\ref{fig:meta.origami2}\textbf{e} compares our predictions for the mean bending angle $\langle \phi \rangle$ with experimental data of the corresponding systems, finding good agreement. Good correspondence is also observed between simulated and experimental standard deviation, $\sigma (\phi)$. It is feasible that the small discrepancies between simulation and experiments emerge from inaccuracies in the WLC model of the springs. Indeed, such a model may be inappropriate for some of the the shorter sections used here (down to 11 nucleotides), especially given that sequence has been ignored. Additionally, the use of WLC springs neglects possible stacking effects at the attachment points of the ssDNA springs with the 16 helix bundles, on either side of the joint. Nevertheless, despite small discrepancies, the automated reconstruction of accurate profiles of mean bending angle (to within 10$^\circ$) confirms the applicability of this method to the rapid prediction of the mechanical and structural properties of DNA origami before manufacture.\\

\section{Conclusions}

Molecular simulation is essential in the design and interpretation of systems which use DNA to build mechanical structures. However, unbiased MD simulation gives little information about the mechanical response of these structures to an applied force. Additionally, for multistable systems with non-ergodic dynamics, unbiased simulation may entirely miss certain states, which may be critical to the function of the construct. To address these limitations, here we have combined well-tempered metadynamics and the popular oxDNA force field, thus introducing a tool for the fast and automated reconstruction of one and two-dimensional free energy landscapes of deformable DNA nanostructures, including sampling of multiple minima and transition states in multi-stable devices.\\

To exemplify the utility of our metadynamics implementation in DNA nanotechnology, we have applied it to four case studies, associated to systems of different scale and conformational complexity. First, we have demonstrated automated sampling of the reversible kinking of a short DNA duplex under compression, replicating experimental and computational observations on the process~\cite{Fields2013,Du2008,Pyne2021,Lankas2006,Harrison2015,Harrison2015a}.
\\We have then reconstructed the free energy landscape of bistable DNA systems whose dynamics would be non-ergodic under conventional MD, even using coarse-grained models. In particular, we have analysed a bistable Holliday junction exhibiting two possible conformers and found remarkable agreement between simulated and experimental conformer occupancy\cite{Joo2004}. The obtained free energy profiles also offered new insights on the effect of ionic strength on the accessibility of the transition state, which qualitatively correlate withe experimental trends in switching rates\cite{Joo2004}. Additionally, we have reconstructed the free energy landscape of a bistable motif previously used for information relaying in DNA origami~\cite{Song2017}, for which we have identified plausible reaction intermediates -- a useful insight for integrating these units into signal transduction architectures.\\
To demonstrate the applicability of our oxDNA MetaD implementation to larger constructs, we have predicted the mechanical response of a compliant DNA origami joint to varying force. We have further shown how, thanks to its ability to map out thermally unaccessible conformational landscapes, MetaD unlocks a new pipeline for the computer-assisted design of joints with prescribed equilibrium angles and stiffness, which we have benchmarked against experimental data~\cite{Zhou2014}. \\

By combining oxDNA with metadynamics, we have enabled faster prediction of free energy profiles without compromising the detail of the underlying DNA model. The process of landscape acquisition can be fully automated as it does not require manual tuning of biasing weights and uses a single thermodynamic window, { contrary to umbrella sampling,}  and can therefore be accessed by users lacking advanced computational expertise. Furthermore, our approach efficiently exploits parallelisation between multiple { CPUs or GPUs.}\\

Our simulation strategy offers a much needed design and characterization tool for the growing community interested in applying DNA nanotechnology to engineer dynamic, reconfigurable devices\cite{Song2017, Li2021}, and nanorobots\cite{Huang2020}, both of which would benefit from rapid \emph{in silico} prediction of free energy landscapes. This is especially the case for large origami structures, composed of multiple DNA scaffolds\cite{Huang2020}, where conventional MD simulation is even more costly. Our technique would also be particularly suited for the better and faster calibration of nanoscopic mechanical probes\cite{Nickels2016,Dutta2018,Stephanopoulos2021}, especially in cases where simple analytical models may yield inaccurate results\cite{Engel2020}. { Finally, MetaD is not only relevant when exploring deformation in fully hydrogen-bonded motifs, but could be also applied to free-energy landscapes associated with hybridization / de-hybridisation by defining suitable collective variables, \emph{e.g.} in terms of number of hydrogen-bonded nucleotides in the system~\cite{Srinivas2013,Clowsley2020}.} In general our approach will enable faster and more detailed acquisition of information related to the mechanical behavior of nanostructures, improving the feasibility of simulation-informed design, and facilitating direct comparison of molecular modelling to experimental measurements.

\section{Methods}

\subsection{oxDNA implementation}

The oxDNA stand-alone executable was extended to enable support for tabulated potentials and corresponding forces between the centres of mass of collections of particles on a one or two dimensional grid (CPU implementation), or a one dimensional grid (CUDA implementation).~\cite{Rovigatti2015} The source code was otherwise unchanged.\\

{ A Python interface was then used to launch multiple MD oxDNA simulations (replicas) in parallel, analyse distributions of collective variables, and update the bias.
Each of the  $N$ replicas was initialised form a different location in collective variable space and simulated under the effect of the time-evolving bias $B_t$, shared between all replicas. After each MetaD cycle, corresponding to $\tau$ MD time-steps, the bias was updated with $N$ Gaussians placed at the instantaneous locations of each of the replicas in configuration space, as discussed above. The parallel replicas therefore share their history to construct an optimal potential, which leads to more efficient exploration of the configurational space and to an $N$-fold speed up in bias convergence\cite{Raiteri2006}. It should also be noted that at early times, the replicas repel each other, encouraging them to search different regions. However, this effect diminishes at later times. Parallel simulations were run on CPUs for dsDNA buckling, Holliday junction isomerization and  bistable unit isomerization, while GPUs were used for the origami compliant joint. The number of replicas used in each case study is reported in table S2.}\\

Following convention\cite{Bussi2020}, the bias was defined on a grid, necessary to avoid slowdown as the number of forces involved increases. The grid spacing, $\delta x$, has value chosen to be at maximum one fifth of the MetaD $\sigma$ -- values are given in table S1. Our implementation is compatible both with Monte Carlo, where the potential felt by the particle is calculated from bilinear interpolation of the gridded bias, and MD, where the force is calculated from the numerical derivative.\\

{ The sequence-averaged version of the oxDNA force field was used in all cases except when mapping the free energy landscape of Holliday junction isomerisation, where the sequence-dependent force field~\cite{Sulc2012} was instead adopted.}\\

{ Simulations for the compliant origami joint required 36 hours over 4 GPUs (total 144 GPU-hours, Nvidia P100 GPU 16GiB), while the other case three studies required between 36 and 60 hours over 32 CPUs (total 1152-1920 CUP-hours, 2$\times$ Intel Xeon Skylake 6142 processors, 2.6GHz 16-core). These timescales represent massive improvements from un-biased simulations which may require tens of thousands of GPU hours for characterising the mechanical behaviour of origami nanomachines~\cite{Sharma2017}.}


{
\subsection*{Choice of MetaD parameters}
The analysis by Laio \emph{et al.}~\cite{Laio2005} and Bussi \emph{et al.}~\cite{Bussi2006} highlighted the influence of MetaD free parameters $\sigma$, $A$, and $\tau$ on the errors of inferred free energies and convergence timescales. These studies recommend optimal choices for the Gaussian width $\sigma$ at a fraction of  the system's size in the collective variable space~\cite{Laio2005}. Because in this work we extract free energy surfaces from configuration sampling of simulations biased with the asymptotic $B_t$, rather than directly from the bias, we { only followed the heuristic consideration that $\sigma$ should be smaller than the lengthscale of the free energy features one wished to map, to prevent over-biasing}. The ratio $A/\tau$ determines the (initial) rate of growth of the bias, and therefore the convergence time, with larger $A/\tau$ implying faster convergence~\cite{Laio2005}. For well-tempered MetaD, $A/\tau$ is not critically important, as the amplitude of corrections decays exponentially. Using small values of $\tau$ (while appropriately re-scaling $A$) reduces ``discreetness" in potential deposition and errors in free energy estimates~\cite{Laio2005}. Because in our implementation the MetaD bias is computed and updated by a Python script, which then re-launches the stand-alone oxDNA executable after each cycle, computational inefficiencies emerge when reducing $\tau$. These were considered in our choices of $\tau$. The MetaD parameters for each of the systems simulated are summarised in table S1.
}

{
\subsection*{Choice of collective variables}
Choice of collective variables in MetaD should follow key criteria, detailed in Bussi \emph{et al.}~\cite{Bussi2020}. First, the collective variables should be designed to force the system to explore the high-free energy transition states one wishes to sample, which is done by ensuring that these states correspond to unique values of the collective variables which are not accessible when the system occupies low-free energy configurations. The application of this criterion is well exemplified by the definitions of the two-dimensional collective variables for our Holliday junction and bistable unit case studies, where the two isomers are clearly separated from the intermediate transition states in the $(x_1,x_2)$ planes. Second, and critical when mapping deformation free energy of large DNA nanostructures, one must ensure that the collective variables are properly coupled to the deformation modes one wishes to characterise. For example, if one would like to study bending of helices or bundles, the collective variables should be defined based on the coordinates of multiple nucleotides on different strands, to avoid that bias buildup leads to rupture of hydrogen bonds and nanostructure disassembly rather than bending.
}

\subsection*{Case studies}
Unless otherwise stated, simulations used the oxDNA2 force field with 0.5 M  ionic strenght, and sequence averaged parameters. Molecular Dynamics was used to sample configurations. A timestep of 0.004 simulation units was used, except for the origami simulation which used a timestep of 0.005 simulation units. To maintain a temperature of {$T=300$ K}, an Andersen-like thermostat was used -- time evolution is Newtonian but every 103 timesteps, a fraction of particles have their velocities drawn from a Maxwell-Boltzmann distribution. The fraction corresponds to a diffusion coefficient of 2.5 oxDNA units. Configurations were saved every $1\times10^5$ timesteps for all systems except the origami, where they were saved every $1\times10^4$ timesteps. Well-tempered metadynamics simulations were run with multiple walkers with parameters listed in tables S1 and S2. Subsequently, the converged bias from those simulations was used in MD to verify correct convergence. Details concerning replicas and timescales are given in table S2.

\subsubsection*{Kink induced buckling in dsDNA}
A DNA duplex of length 30, with sequence 5'-ATG CAC AGA TTA GGA CCA ACC AGG ATA GTA-3' was initialized using the generate-sa.py script in the oxDNA software package. MetaD was run with a bias on the collective variable $x$, the distance between virtual particles at the centres of mass of the six nucleotides on one end of the duplex and the corresponding six at the other end. { This choice was made to guarantee that the applied bias induces duplex bending, rather than de-hybridization.} Details of simulations are given in tables S1 and S2.\\

To evaluate a reference free energy -- $\Delta G(x)$ -- the bias from the $\Delta T=16 T$ system was used in biased MD to acquire a large number of states (table S2). Convergence of the free energy implied by the bias to the reference free energy is demonstrated in figure S2.  Here, an equilibration period of $1\times10^8$ timesteps was used to decorrelate initial states. To demonstrate convergence, the $\Delta G(x)$ values were constructed from either the first half or the second half of the simulation, see figure S2\textbf{a}. Differences between the two are substantially smaller that the width of lines used to plot.\\

{Figure \ref{fig:meta.dsDNA}\textbf{f}} features a two dimensional free energy landscape. The quantity on the $y$-axis, $U_\text{4th least stacked}$, was chosen to distinguish the buckled from the unbuckled state. This energy is defined by first acquiring the 5' and 3' stacking energies associated with each non-terminal nucleotide. Subsequently, the lesser of these two values was stored for each nucleotide. The fourth greatest (\emph{i.e.} least negative) value in the list then defined $U_\text{4th least stacked}$. Since the buckled state breaks two internal base pair stacking interactions (where each is between a pair of nucleotides), this value rises to $0$ if the duplex is buckled.\\

To illustrate the two distinct buckled and unbuckled states we have plotted a kernel density estimator (KDE) with bandwidth 0.05 units -- either nm or $k_BT$ in figure \ref{fig:meta.dsDNA}\textbf{b}. This should not be overinterpreted other than to imply bistability when $x$ is constrained to a value below $\approx 6$\,{nm}. For example, the buckled state has $U_\text{4th least stacked}$ exactly zero, so the density here is very high, and the exact free energy values will depend strongly on KDE bandwidth.\\

\subsubsection*{Holliday junction isomerization}

Holliday junctions were based on the J3 junction, as previously studied using single molecule FRET measurements\cite{Joo2004}, and an alternative coarse-grained force field\cite{Wang2016}. Here, the junction is truncated so that arms are each 12 bps or $\approx4$\,nm long, slightly over 4 Debye lengths for 100\,mM ionic strength; sequences are given in table S3. Truncation was necessary for faster simulation, and it is unlikely that nucleotides so far from the junction contribute to configuration probabilities. Structures were initialized using the MrDNA \cite{Maffeo2020} software, then subsequently refined in the oxDNA-viewer software \cite{Poppleton2020}. Four sets of simulations were run: either with a sequence averaged force field at 500\,mM ionic strength, or with a sequence specific force~\cite{Sulc2012} field at either 500\,mM, 200\,mM, or 100\,mM ionic strength. In each case, the counterion is modelled implicitly through control of the Debye length over which electrostatic screening operates. As simulations at reduced electrostatic screening are slower, runs at 100\,mM ionic strength necessarily have fewer steps.\\

A two dimensional collective variable was used in MetaD simulation, $(x_1,x_2)$, as illustrated in figure \ref{fig:meta.nanostar}\textbf{a}. { These variables were designed to clearly distinguish the two stacked isomers, where $x_{1/2}$ take small values and   $x_{2/1}$ large values, from the un-stacked transition state where $x_1$ and $x_2$ both have high values, incompatible with the stacked isomers.}
Simulations were performed with the parameters from table S1.  After MetaD runs, biased MD simulation were initialized from the terminal states of each of the six metadynamics walkers, each in eight replicas. The first $1\times10^7$ steps were discarded to allow for decorrelation. Contour plots in figure \ref{fig:meta.nanostar}\textbf{c} are acquired from histograms of biased MD. 
Convergence of the free energy implied by the MetaD bias and comparison to that acquired by histograms of biased MD is illustrated in figure S3.\\

For identification of the states \emph{iso}I, \emph{iso}II, and the intermediate, the following criteria were used. For both the stacked-X conformers and the intermediate we required that all hydrogen bonds in the eight nucleotides adjacent to the junction were formed (internal energy $<-1\,k_B T$). For the stacked-X conformers, we further required that stacks were formed between pairs of neighbouring ``arms" at the junction, with a stack being said to occur if its internal energy is $<-5\,k_B T$.   We thus identified the \emph{iso}I and \emph{iso}II conformers based on which stacks were formed. As expected, stacked-X states, with two formed and two un-formed stacks as illustrated in figure \ref{fig:meta.nanostar}\textbf{a}, are dominant in all explored conditions. The intermediate was defined as the state with no stacks formed, but all hydrogen bonds present. The aforementioned state definitions were used to acquire the probabilities and free energies in figure \ref{fig:meta.nanostar}\textbf{f}. The \emph{iso}I and \emph{iso}II states for the sequence averaged force field should be equal by symmetry, so the $\approx 3$\% difference in state probability is a reasonable indication of simulation convergence. Errors of estimates are given as one standard error, using 6 repeats initialized from different positions in the collective variable space.\\


\subsubsection*{Bistable unit isomerization}

The bistable unit studied in Supplementary Discussion 2 was designed in caDNAno \cite{Douglas2009}. The strand routing is given in figure S4.  A two dimensional order parameter was used to bias MetaD, based on distances $x_1$ and $x_2$ as illustrated in figure S5\textbf{a}. These distances were defined between groups of four nucleotides adjacent to each of the four nicks in the structure, { making sure that the two isomers are clearly separated form the intermediate transition state on the $(x_1, x_2)$ plane, as done for the case of the Holliday junction.} \\

Parameters for simulations are listed in  table~S1, with convergence illustrated by the plots in figure~S6. The terminal states of six walkers were then used as initial states in  MD simulations, biased with the converged $B_t$ from MetaD. The biased MD simulations were used to construct the free energy distribution in figure S5\textbf{b} (left) ($3\times10^7$ steps discarded prior to collection for decorrelation). \\

Additionally, MetaD was run with a 1D collective variable, $\arctan\frac{x_2}{x_1}$. To use this order parameter, analytical derivatives with respect to position were calculated. MetaD simulations were run with parameters listed in table~S1, with duration listed in table S2. After convergence of the 1D bias, multiple MD simulations were then run with said bias, where parameters for runs have been listed in table S2. These biased MD runs were used to reconstruct the 2D free energy landscape given in figure S5\textbf{b} (right).\\

\subsubsection*{DNA origami compliant joint bending}

The DNA origami compliant joint studied here was based on an experimentally realised structure \cite{Zhou2014}. However, for reasons of speed, it was truncated, reducing the length of the helix bundles on the two sides of the joint. The experimental structure had six ssDNA scaffold sections routed across the compliant joint to apply a bending moment, whose magnitude could be controlled by the ssDNA length. Here we have removed these sections, relying instead on the MetaD bias to bend the joint. The caDNAno routing is given in figure S7. 

A collective variable was defined as detailed in figure \ref{fig:meta.origami1}\textbf{a} and discussed in the main text. After design in caDNAno, structures were relaxed \emph{via} a gradient descent to prevent large forces. Simulations were then run with the GPU-accelerated version of oxDNA~\cite{Rovigatti2015}. For metadynamics, four walkers were run in parallel on separate GPUs associated with the same compute node. MetaD parameters were used as detailed in table S1.\\

To establish a reference free energy ($\Delta G$) to validate the convergence of MetaD predictions and later evaluate the $\phi$ distribution, the MetaD bias was frozen, and biased MD simulations were run. As detailed in table S2, four different initial $x$ configurations were used to generate samples, with six replicas run from each of those four initial configurations. These were run for 4$\times 10^6$ steps to decorrelate replicas, followed by a production run (table S2 for details). To evaluate uncertainties for all estimates, the standard error from simulation runs initialized from different initial configurations was used.\\

To evaluate the distribution of $\phi$, the reference beads illustrated in figure \ref{fig:meta.origami1}\textbf{a} (yellow beads) were used. The top and bottom sections of the bundle each have 12 reference nucleotides selected. These are are composed of two groups of six, one further and one nearer to the joint. Each of those groups of six corresponds to three base pairs, chosen to be adjacent to crossovers, { guaranteeing that the applied bias does not induce unwanted structure disassembly}. The distance along the bundle between the near six and far six was chosen to be 21 nucleotides (two helical turns), so that base pairs used as references have the same orientation. The centre of mass of each of the four groups was acquired. For convenience, we use the notation $\vec{x}_\text{near}^\text{\: top}$, $\vec{x}_\text{far}^\text{\: top}$, $\vec{x}_\text{near}^\text{\: bottom}$, $\vec{x}_\text{far}^\text{\: bottom}$ to denote these centres of mass. Every 20,000 steps, the locations of the centres of mass was saved. Subsequently, two vectors were defined:

\begin{align}
    \vec{v}^{\:\text{top}}&=\vec{x}_\text{far}^{\:\text{top}} - \vec{x}_\text{near}^{\:\text{top}},\\
    \vec{v}^{\:\text{bottom}}&=\vec{x}_\text{far}^{\:\text{ bottom}} - \vec{x}_\text{near}^{\:\text{bottom}}.\\
\end{align}

The angle between these two vectors was used to define $\phi^{\text{(0,180)}}$. This angle is not the $\phi$ that is then used in free energy calculations. It is important to then consider a definition of $\phi$ on $(0^\circ,360^\circ)$, rather than $(0^\circ,180^\circ)$ (so that in figure \ref{fig:meta.origami1}\textbf{c}, location E, corresponds to a $\phi>180^\circ$, while figure \ref{fig:meta.origami1}, location F, corresponds to a $\phi<180^\circ$). Therefore, an additional vector was defined, corresponding to the direction into the page in figure \ref{fig:meta.origami1}\textbf{a} (far left). We label this $\vec{v}_\text{ortho}$, and defined it using the two sets of beads furthest from the location of the junction. Looking at figure \ref{fig:meta.origami1}\textbf{a} (far left), $\vec{v}_\text{ortho}$ corresponds to the average vector from the centre of mass of the yellow beads nearest the reader to those into the page. Subsequently, the value of $${\text{sign}(( \vec{v}^\text{bottom} \wedge  \vec{v}^\text{top})  \cdot \vec{v}_\text{ortho})}$$ was acquired. This takes a value which is negative if $\phi<180^\circ$, and positive otherwise. Hence $\phi$ was acquired as:

\begin{align}
    \begin{aligned}
    \phi= 
        \begin{cases}
            \phi^{\text{(0,180)}}, \text{if } ( \vec{v}^{\:\text{bottom}} \wedge & \vec{v}^{\:\text{top}})  \cdot \vec{v}_\text{ortho}\\&<0 \\
            360-\phi^{\text{(0,180)}}              & \text{otherwise}.\\
        \end{cases}
    \end{aligned}
\end{align}

To evaluate the effect of ssDNA springs on the bending angle, the following approximations were used. There are two sets of three ssDNA which bridge the compliant DNA origami joint gap in the experimental system. These correspond to one set of three ssDNA segments which bridge the short gap, and one set which bridge the long gap (where the short and long gaps are illustrated in {figure \ref{fig:meta.origami2}\textbf{c})}. The set of three ssDNA sections which bridge the short gap each have $N_\text{short}$ ssDNA nucleotides; the others have $N_\text{long}$ ssDNA nucleotides.

We have then evaluated the free energy contribution from each of the ssDNA springs, {$\Delta G_\mathrm{WLC}^{N_\text{nts}}(L)$}, using an analytical approximation for the free energy of a WLC:\cite{Petrosyan2017} 

\begin{align} 
\begin{aligned}
   { \Delta G_\mathrm{WLC}^{N_\text{ nts}}(L)} = \frac{k_BT}{L_p} \int_0^L dx\bigg( &\frac{1}{4}\big( 1-\frac{x'}{L_0}\big)^{-2} \\
        &- \frac{1}{4} + \frac{x'}{L_0} \\
        &- 0.8\big( \frac{x'}{L_0}\big)^{2.15}\bigg).
\end{aligned}
\end{align}

Here $L_p$ is the persistence length of ssDNA, which we have taken as 2 nm\cite{Roth2018}, while $L_0$ is the contour length. This was acquired from assuming that the contour length of ssDNA was 0.676 nm/nt\cite{chi2013}. One subtlety is that the number of nucleotides $N_\text{nts}$ refers to is one greater than the number in the actual chain. This may seem surprising, but consider that the case where there are $0$ nucleotides in the ssDNA spring; there would still be 1 nucleotide of separation between the two sides of the joint. To evaluate the total free energy, we summed the contributions from the six chains, three of which contain $N_\text{short}$ nucleotides, and three of which contain $N_\text{long}$ nucleotides. 

There are six springs in total, so the total statistical weight used to compute averages is:

\begin{align}
\begin{aligned}
    \exp\big(\beta({B_t(x)}-&\sum_{i=0}^2 {\Delta G_\mathrm{WLC}^{(N_\text{short}+1)}(L_i)} \\
    - &\sum_{i=3}^5 {\Delta G_\mathrm{WLC}^{(N_\text{long}+1)}(L_i)})\big).
\end{aligned}
\end{align}

Here {$B_t(x)$ is the MetaD bias}, $i\in\{0,1,2\}$ indexes short springs, while $i\in\{3,4,5\}$ indexes long springs. For each, $L_i$ is the separation distance measured in simulation between attachment points. This weighted distribution was used to acquire both $\langle\phi\rangle$ and $\sigma(\phi)$, as plotted in { figure \ref{fig:meta.origami2}\textbf{e}-\textbf{f}}. Uncertainties here were acquired from the standard error over repeating this procedure for simulations run with four different initial conditions uniformly spaced in the range of $x$ studied.\\





\FloatBarrier

\begin{acknowledgement}

LDM acknowledges support from a Royal Society University Research Fellowship (UF160152) and from the European Research Council (ERC) under the Horizon 2020 Research and Innovation Programme (ERC-STG No 851667 -- NANOCELL). W.T.K. acknowledges funding from an EPSRC DTP studentship. This work was performed using resources provided by CSD3 operated by the University of Cambridge Research Computing Service (www.csd3.cam.ac.uk), provided by Dell EMC and Intel using Tier-2 funding from the EPSRC (capital grant EP/P020259/1), and DiRAC funding from STFC (www.dirac.ac.uk).\\
{ The developed simulation code is available at https://doi.org/10.5281/zenodo.6326800, along with examples based on the case-studies presented here}.

\end{acknowledgement}

\begin{suppinfo}

Supplementary Discussions, Figures and Tables.

\end{suppinfo}

\bibliography{biblio}

\end{document}






\clearpage

\section*{Supplementary Notes}
\begin{figure*}[ht!] 
\centering    
\includegraphics[width = 160mm]{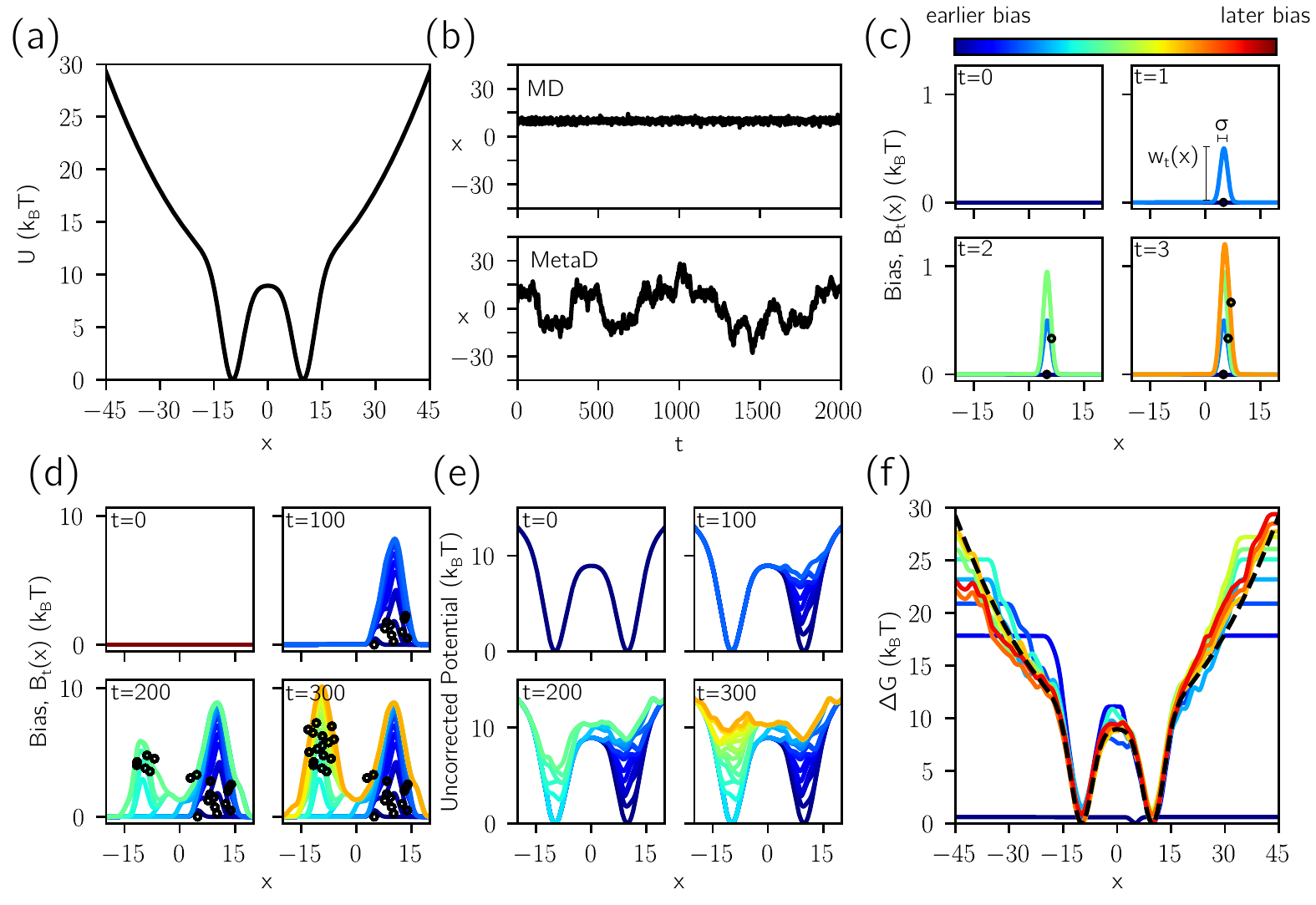}
\caption[A 1D landscape illustrating the principle of metadynamics]{A 1D landscape illustrating the principle of metadynamics. \textbf{(a)},~Consider a particle moving with stochastic dynamics on the 1D potential energy landscape illustrated. Here, two potential energy minima are separated by a $\approx 10\: k_BT$ transition state. \textbf{(b)},~The dynamics of MD on this landscape are non-ergodic; after initialisation in one potential energy minimum, the particle cannot escape in the timescale of simulation.  By contrast, MetaD learns to escape the first minimum after 200 iterations of metadynamics. (Note that for consistency with the literature, $t$ refers to the number of iterations of metadynamics, not the number of timesteps.)  \textbf{(c)},~Metadynamics learns a history dependent potential, $B_t(x)$, whose update is illustrated here. The bias is initialised to 0 ($t=0$). The position of the particle is illustrated by the black circle. A repulsive Gaussian potential is added at $t=1$; this has height $w_t(x)$ and standard deviation $\sigma$. The system is then evolved under the action of both the potential, $U(x)$, and the bias, $B_t(x)$.  \textbf{(d)},~Bias evolution at longer timescales: black circles indicate positions, where the higher the point, the more recent. \textbf{(e)}, The uncorrected (residual) potential felt by the system: ${U(x)+B_t(x)}$. Although initially the uncorrected potential is bistable, at $t=300$ it has largely been corrected to an almost flat landscape. \textbf{(f)}, The free energy ($U$ for this trivial particle) can be estimated by $\frac{\Delta T + T}{\Delta T}B_t(x)$ after the bias converges. The true landscape here is illustrated by a dashed black line, and the time evolving free energy implied by the bias by coloured lines.
}
\label{fig:meta.theory}
\end{figure*}

{
\subsection*{Supplementary Note 1: MetaD applied to a simple 1D walker}

To illustrate the principle of metadynamics here we consider a 1D walker of unit mass moving with Langevin dynamics in a bistable potential as shown in figure~\ref{fig:meta.theory}\textbf{a}. For convenience, $k_BT$ is set to 1. The underlying potential energy landscape is:

\begin{align}   
    \begin{aligned}
    U(x) = &-A_1\cdot e^{-(x-x_1)^2/2\sigma^2} - A_2\cdot e^{-(x-x_2)^2/2\sigma^2} \\
    &+ A_3\cdot x^2.
    \end{aligned}
\end{align}

Here, $A_1=A_2=10$, $A_3=0.01$, $x_1=-10$, $x_2=10$, $\sigma=3$.

Langevin dynamics were evaluated by Euler integration\cite{Kloeden1992}:

\begin{align}
\begin{aligned}
    x_{\upsilon+\Delta \upsilon} &= x_\upsilon+\dot{x}_{\tau}\: \Delta \upsilon\\
    \dot{x}_{\upsilon+\Delta \upsilon} &= -\gamma\:\dot{x}_\upsilon\cdot \Delta \upsilon
                &+ \sqrt{2\gamma}\: \Delta W_\upsilon- \frac{dU(x)}{dx}\: \Delta \upsilon.
\end{aligned}
\end{align}

Here $\Delta \upsilon$ is the integration step, while the subscript $\upsilon$ denotes time. The linear damping coefficient is set to $\gamma=10$. $\Delta W_\upsilon$ is a delta-correlated noise, chosen independently at each time step from a normal distribution with variance $\Delta \upsilon$, centre $0$. $U(x)$ and its numerical derivative were evaluated through linear interpolation on a grid. The integrator used a timestep of $\Delta \upsilon=0.01$. The metadynamic parameters were $A=0.5$, $\sigma=1$, $\tau = 1000$, $\Delta T=5$.

Figure \ref{fig:meta.theory}\textbf{b} (top) shows a conventional MD trajectory. A configuration initialised in one minimum cannot cross to the other, as doing so would require passing through an unlikely transition state. The histogram approach of evaluating free energies will therefore only describe a narrow set of conformations as the existence of the unsampled free energy minimum cannot be inferred from the trajectory.  

MetaD flattens the free energy landscape, enabling otherwise unlikely transitions between local minima as illustrated in figure \ref{fig:meta.theory}\textbf{b} (bottom). Although initially the particle is trapped in one well, it subsequently transitions to the other, before starting to reversibly visit both wells with a diffusive motion, unaffected by the potential barrier. This time-dependent shift in dynamics is enabled by an history-dependent bias potential, built as discussed in the main text (equations 1-3).

For our simple example, the time evolution of the learned bias is illustrated in figures \ref{fig:meta.theory}\textbf{c},\textbf{d}. Initially, the system is trapped in one minimum (figure \ref{fig:meta.theory}\textbf{d}, t=100), but it builds up a bias which allows exploration of the potential energy well. Eventually the potential landscape experienced by the particle (from combining bias and true free energy) no longer prevents access of the transition state.

Equation 3 in the main text shows how, for well-tempered MetaD, $B_t$ converges to a fraction of the true free energy. For an illustration of this convergence, consider the time evolution of the uncorrected potential, ${B_t + U}$, in our simple example (figure \ref{fig:meta.theory}\textbf{e}). Although initially, the uncorrected potential is $U$, it subsequently becomes flattened by the bias, converging to ${\frac{T}{T+\Delta T} U}$ (cnf main text, noting that in this 1D example $U \equiv \Delta G$).  Analogously, figure \ref{fig:meta.theory}\textbf{f} shows convergence of $\frac{\Delta T + T}{\Delta T}B_t$ to $U$. MetaD therefore provides a method to acquire free energy profiles from the converged bias, even in systems with high free energy transition states.}

\clearpage

\subsection*{Supplementary Note 2: Isomerisation in a re-configurable bistable tile}




Base stacking interactions have previously been used to create reconfigurable devices \cite{Gerling2015}, and a similar approach has recently been applied to design origami-based reconfigurable molecular arrays, which can spatially relay information through the propagation of conformational transitions \cite{Song2017}. The elementary unit of the array is a bistable motif, in which transition between the two configurations requires stacking interactions to break. These elements can be tessellated, creating a rectangular array that preserves the bistability of the individual unit. As a further case-study, here we use our oxDNA MetaD approach to map the bistable free energy landscape of this elementary motif.

\begin{figure*}[p] 
\centering    
\includegraphics[width=\textwidth]{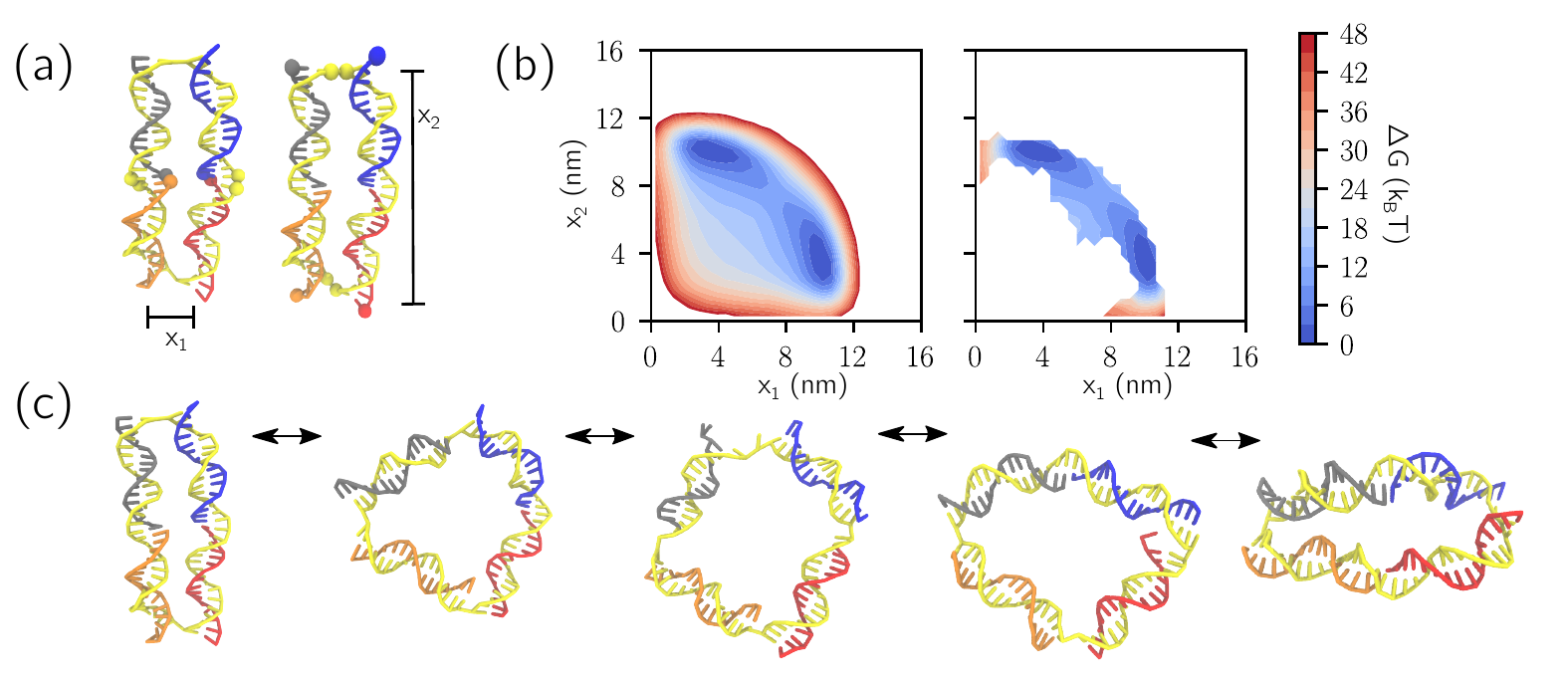}
\caption[A bistable unit undergoes isomerisation]{MetaD enables mapping of the free energy landscape of a switchable bistable unit. \textbf{(a)},~A snapshot of a likely conformation of the bistable unit. Collective variables $x_1$ (left), and $x_2$ (right), are distances between centers of mass of the indicated beads. These were used to bias the dynamics and encourage transitions. \textbf{(b)},~(Left) Well-tempered MetaD simulation ($\Delta T=20T$) provides a bias which flattens the free energy landscape. This enables static bias MD to reconstruct the symmetric bistable distribution expected (convergence of the bias is demonstrated in figure S6). (Right) Biasing can instead be done in one dimension to accelerate convergence -- here the angular coordinate $\theta=\tan^{-1}(x_1/x_2)$ is used. This confines the region sampled to a thin shell around the free energy minimum for each value of $\theta$; since a smaller configuration volume is sampled, convergence is faster. \textbf{(c)},~A proposed transition mechanism, observed as the first transition between conformers for both the 1D and 2D metadynamics simulations.
}
\label{fig:meta.bistable}
\end{figure*}

Figure~\ref{fig:meta.bistable}\textbf{a} shows the structure under consideration, which is a slightly truncated version of the unit used in bistable origami \cite{Song2017}. In our implementation, each of the shorter strands is 17 nts long (1.5 helical turns), against 21 nts in the origami (2 helical turns).\cite{Song2017} Additionally, in experiments, the longer strand would have a strand break at some location -- however, to maintain symmetry, it is circular in the studied model. We do not expect either of these features to have a qualitative effect on the free energy landscape.

When simulated under unbiased MD, the structure remains trapped in a single conformer. A two dimensional reaction coordinate is defined corresponding  to the distances $x_1$ and $x_2$  between centers of mass of the four nucleotides adjacent to each strand break, as illustrated in figure \ref{fig:meta.bistable}\textbf{a}. In a single conformer, one of these values takes a higher value (the helical rise of 34 bps $\approx11.5$ nm), while the other takes a lower value (slightly over the width of a helix $\approx 2.1$ nm).

\begin{figure*}[p] 
\centering    
\includegraphics[]{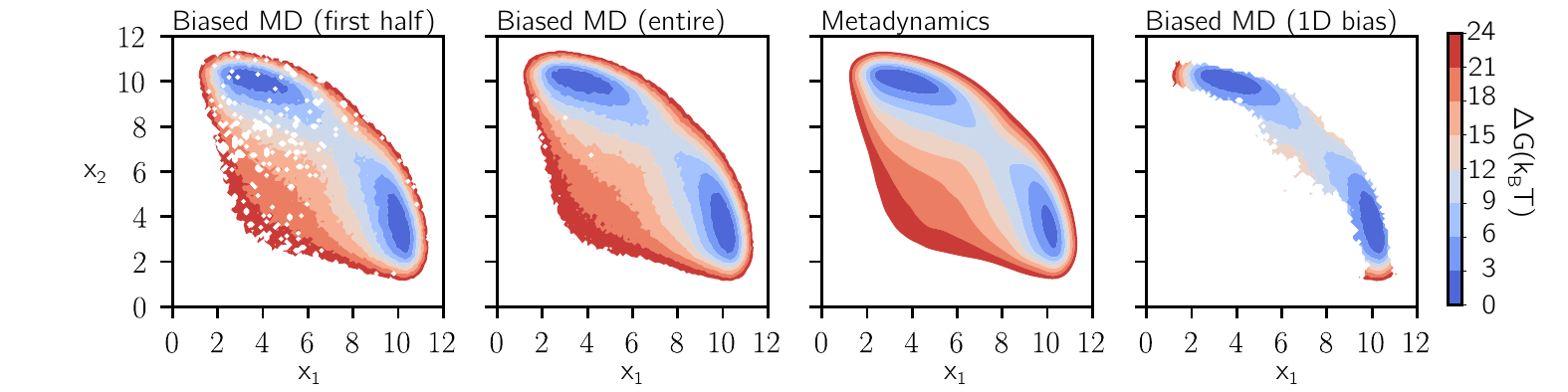}
\caption[Demonstrations of convergence in bistable unit isomerisation]{Demonstrations of convergence in bistable unit isomerisation. The distribution acquired from the first half of biased MD simulation is essentially identical to that from the entire trajectory, bar some sparsity of sampling. This is similar to that acquired by MetaD, although the latter was not run for sufficiently long to correct the asymmetry in the sizes of the two free energy minima -- these should be symmetric due to the symmetry of the molecule.  Running a 1D MetaD simulation based on angular coordinate $\theta = \arctan (x_2/x_1)$ enables acquisition only of likely states as a function of this angle -- hence the limited region of converged observations. Here, we have plotted the distribution from a { MD} run with a static 1D bias chosen from the converged MetaD simulation run. 
}
\label{fig:meta.bistable_convergence}
\end{figure*}

MetaD simulation along this collective variable constructs a bias which enables frequent transitions between the two free energy minima, and can be used to extract the free energy landscape shown in figure \ref{fig:meta.bistable}\textbf{b} (left). Demonstrations of convergence are illustrated in figure~\ref{fig:meta.bistable_convergence}. The symmetry of the free energy landscape structure was not externally imposed, and emerges simply from the multiple transitions between energetically similar conformers.

One of the disadvantages of a 2D collective variable is the significant simulation time necessary for convergence. The structure of the free energy landscape in this example indicates that one could simply bias by the angle subtended by the x-axis and a line from the origin to ($x_1$, $x_2$), \emph{i.e.} $\theta=\arctan\frac{x_2}{x_1}$, as plotted in figure \ref{fig:meta.bistable}\textbf{b}. This enables a faster convergence, and acquisition of the landscape illustrated in figure \ref{fig:meta.bistable}\textbf{b} (right). Here, biasing is only possible in $\theta$, so that investigation into the free energy as a function of $r$ can only be done thermally -- hence the structure revealed here corresponds to a small region in the radial coordinate, $r$, expanded at the minimum for each angular coordinate $\theta$.

The free energy projections in figure \ref{fig:meta.bistable}\textbf{b} look deceptively continuous. However, analysis of actual trajectories indicates that the system undergoes transitions between stacking states identified by their coaxial stacking configurations. Since there are four possible stacking locations, each of which can be either stacked or unstacked, there are a total of 16 states. Geometrical requirements eliminate essentially all of these configurations except those illustrated in figure \ref{fig:meta.bistable}\textbf{c}, which show the transition path followed at early times in both 1D and 2D metadynamics. A single coaxial stack detaches, which is followed by transition into an entirely unstacked intermediate, the formation of one of the stacks in the alternative conformer, and finally that of the opposite stack.

However, we note that great care must be taken in interpreting these transitions as representative of the spontaneous transition path of the system, as they are strongly influenced by the reaction coordinate and the accumulating bias. While looking at early simulation times mitigates against the effect of biasing, accurate assessment of transition mechanisms must be performed using inherently dynamical rare event simulation approaches such as Forward Flux Sampling \cite{Allen2009}.

\clearpage

\section*{Supplementary Figues}

\clearpage

\begin{figure*}[p] 
\centering    
\includegraphics[]{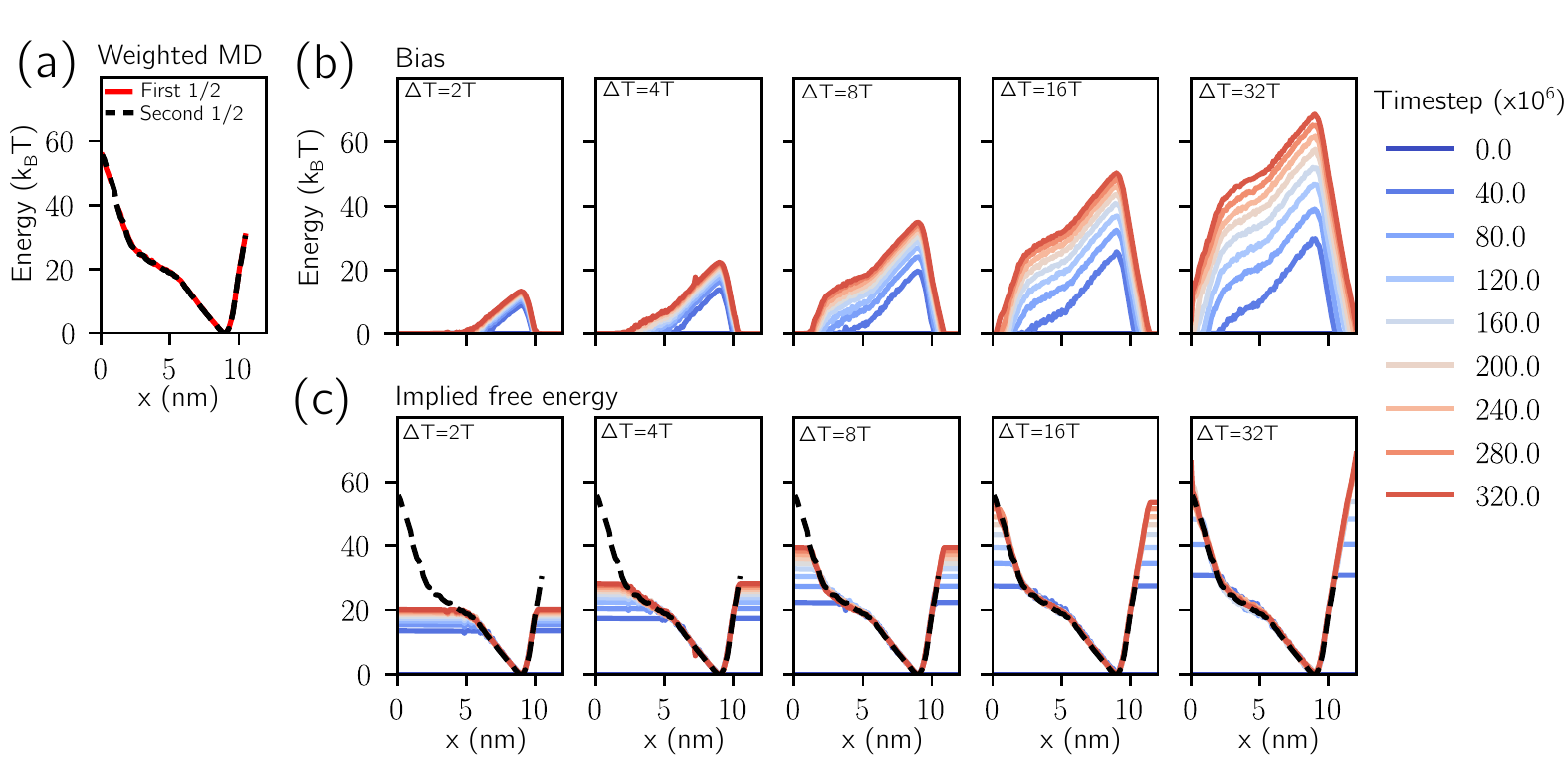}
\caption[Demonstrations of convergence in dsDNA buckling]{Demonstrations of convergence in dsDNA buckling. \textbf{(a)},~the free energy as a function of $x$ from biased MD simulation -- results inferred from the first and second halves of MD simulation are plotted indicating convergence. \textbf{(b)},~time-dependent growth of the bias potential for various values of $\Delta T$. \textbf{(c)},~Free energies implied from MetaD, overlaid with the known free energy from weighted molecular dynamics (black dashed line). { The legend, applying to panels \textbf{c} and \textbf{d} indicates MD time-steps from the start of the simulation}.}
\label{fig:meta.dsDNA_convergence}
\end{figure*}

\clearpage

\begin{figure*}[p] 
\centering    
\includegraphics[]{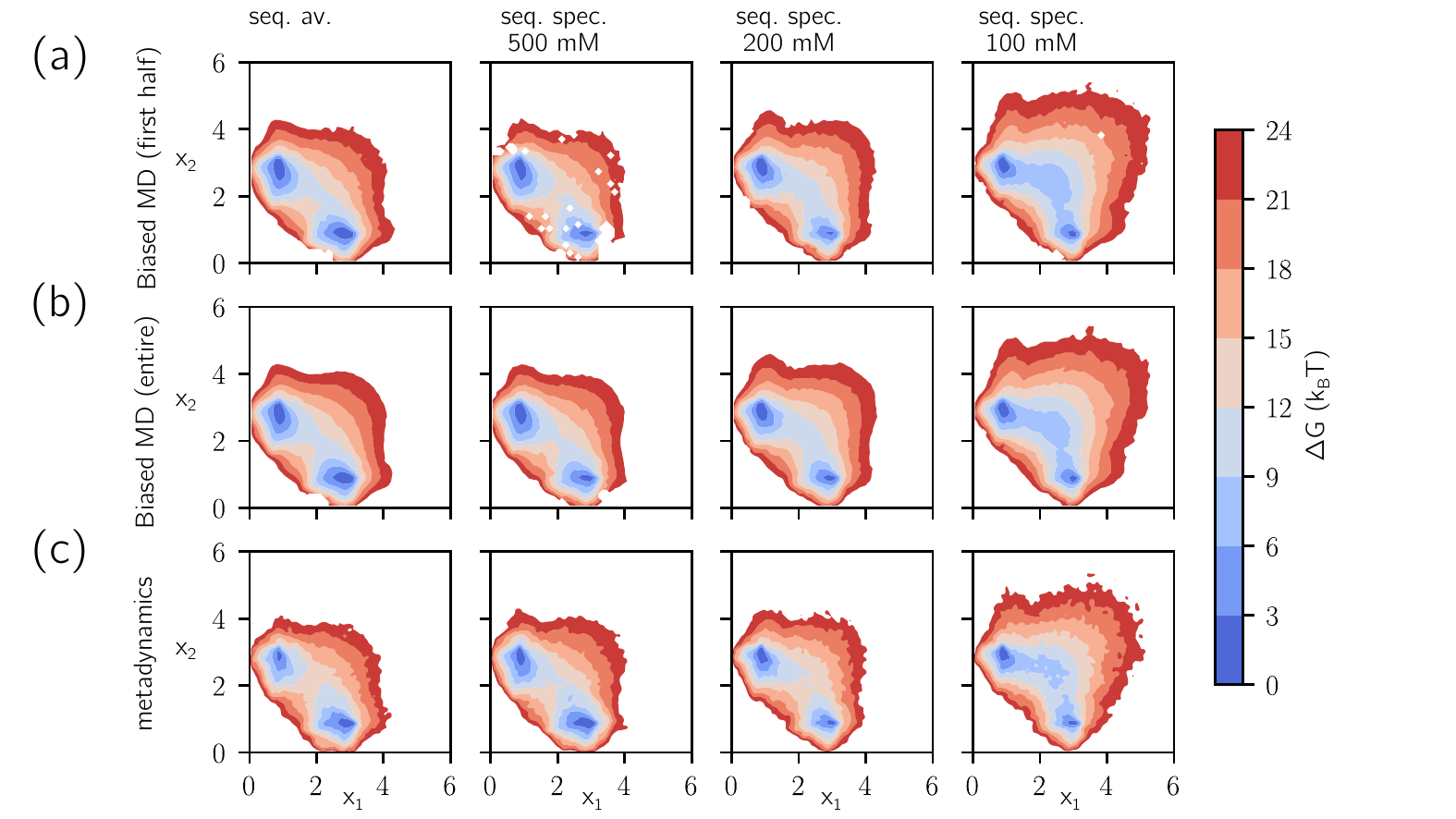}
\caption[Demonstrations of convergence in Holliday junction isomerisation]{Demonstrations of convergence in Holliday junction isomerisation. \textbf{(a)},~Free energy distribution estimated by a histogram of biased MD simulation during the first half of simulation. \textbf{(b)},~Distributions of the same, but over the entire simulation period. \textbf{(c)},~Implied free energies from metadynamics simulation. The slightly speckled structure is a consequence of the small $\sigma$ used in metadynamics. 
}
\label{fig:meta.nanostar_convergence}
\end{figure*}

\clearpage

\begin{figure*}[p]
\begin{center}
\includegraphics[width=\textwidth]{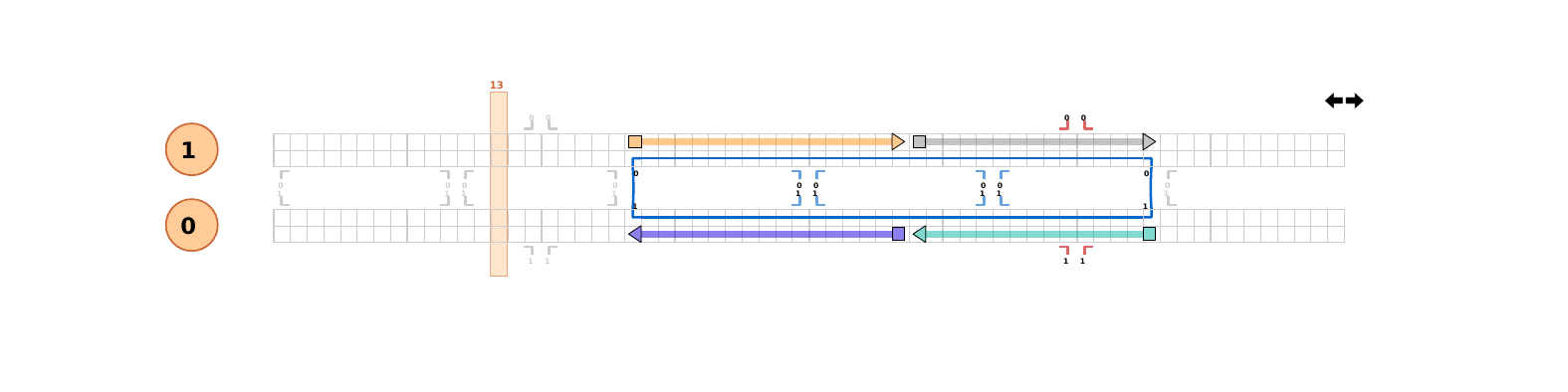}
\end{center}
\caption[caDNAno routing of the bistable unit]{caDNAno routing of the bistable unit.
}
\label{fig:appendix.bistable}
\end{figure*}

\clearpage

\clearpage

\clearpage

\begin{figure*}[p]
\begin{center}
\includegraphics[width=\textwidth]{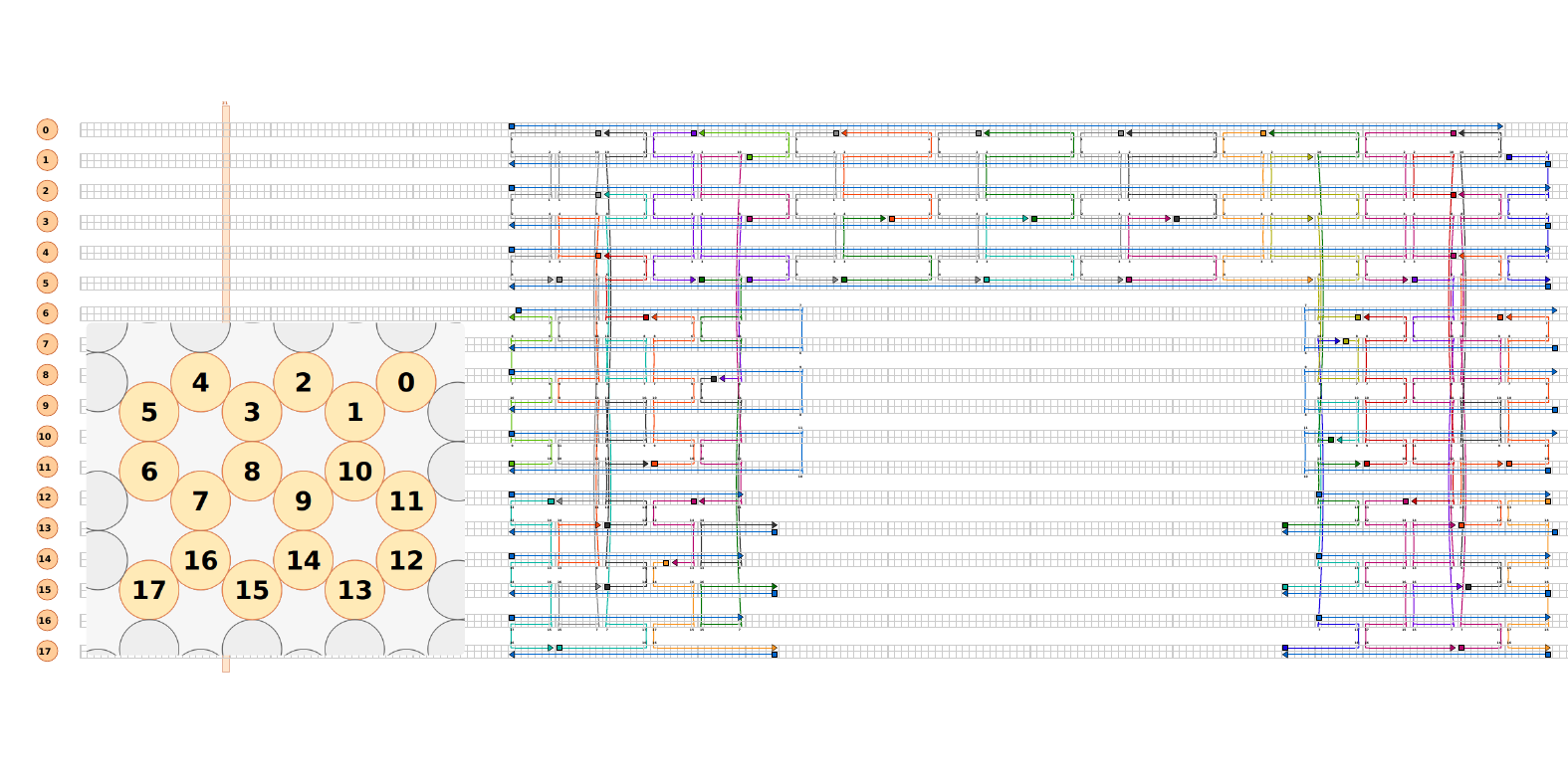}
\end{center}
\caption[caDNAno routing of the truncated compliant joint]{caDNAno routing of the truncated compliant joint. 
}
\label{fig:appendix.compliant}
\end{figure*}

\clearpage

\section*{Supplementary Tables}


\begin{table}[]
\caption[Parameters for metadynamics simulation]{Parameters for MetaD simulations. { $X_\text{min}$ and $X_\text{max}$ indicate the limits of the range accessible to the collective variables.} The meaning of all other parameters is described in the main text. ``ox." indicates to oxDNA units.}
\label{tab:metadynamics.MetaD}
\begin{tabular}{@{}lllllllll@{}}
System                & A ($k_BT$) & $\sigma$ (ox.) & $\tau$ (steps) & {$\Delta T/T$}  & $X_{\text{min}}$ (ox.) & $X_{\text{max}}$ (ox.) & $\delta x$ \\
dsDNA        & 0.5      & 0.05           & $4\times10^5$  & 2,4,8,16,32 & 0                      & 20                     & 0.005         \\
Bistable    & 0.5      & 0.5            & $4\times 10^5$ & 20          & -1                     & 30                     & 0.1           \\
Bistable (1D) & 0.5      & 0.01           & $2\times 10^4$ & 20          & -1                     & 1.7                    & 0.002           \\
HJ              & 1        & 0.1            & $1\times10^5$  & 20          & -0.1                   & 10                     & 0.005    \\
Compliant joint              & 1        & 2           & $1\times10^5$  & 32          & 0                   & 80                     & 0.1         
\end{tabular}%
\end{table}

\vskip 1cm
\begin{table}[]
\caption[Timesteps and replicas for simulation]{Numbers of timesteps and replicas for MetaD simulation and biased MD simulation.}
\label{tab:metadynamics.times}
\begin{tabular}{@{}lllll@{}}
System                & $N_\text{steps}$ (MetaD) & $N_\text{replicas}$ (MetaD) & $N_\text{steps}$ (MD) & $N_\text{replicas}$ (MD) \\
dsDNA        & $300\times10^6$         & 32                         & $1\times10^9$         & 105                      \\
Bistable    & $200\times10^6$         & 32                         & $230\times10^6$       & 48                       \\
Bistable (1D) & $190\times10^6$         & 6                          & $190\times10^6$       & 48                       \\
HJ (seq. av.)   & $355\times10^6$         & 6                          & $500\times10^6$       & 48                       \\
HJ (500 mM)     & $260\times10^6$         & 6                          & $500\times10^6$       & 48                       \\
HJ (200 mM)     & $355\times10^6$         & 6                          & $400\times10^6$       & 48                       \\
HJ (100 mM)     & $280\times10^6$         & 6                          & $250\times10^6$       & 48                      \\
Compliant joint     & $3\times10^7$         & 4                          & $1\times10^7$       & 24                      
\end{tabular}%
\end{table}

\vskip 1cm

\begin{table}[h!]
\caption{Sequences of the truncated J3 Holliday junction (5' $\rightarrow$ 3').}
\begin{center}
    \begin{xltabular}{\textwidth}{@{}lX@{}}
    \textbf{Name}             & \textbf{Sequence}                                 \\
    \endhead
    \textbf{A}  &  CGGTAGCAGCC TGAGCGGTGGT\\
    \textbf{B}  &  ACCACCGCTCA ACTCAACTGCA\\
    \textbf{C}  &  TCCTAGCAAGG GGCTGCTACCG \\
    \textbf{D}  &  TGCAGTTGAGT CCTTGCTAGGA\\
    \end{xltabular}
\end{center}
\end{table}

\clearpage

\bibliography{biblio}